\documentclass[
               DIV=15, font size=10.5pt,
                 onecolumn,
               ]{scrartcl}  

\usepackage[dvipsnames]{xcolor}
\usepackage{psfrag} 
\usepackage{subfig}
\usepackage{hyperref}
\usepackage{amsmath}
\usepackage{graphics}
\usepackage{mathtools}
 \usepackage{pstool} 
\usepackage{todonotes}
\usepackage{gensymb}
\usepackage{bm}
\usepackage{multirow}
\usepackage{cancel}
\usepackage{float}
\usepackage{textcomp}
          
\usepackage[square, numbers,sort&compress]{natbib}
 
\usepackage{pgfplots}
\usepackage{pgfplotstable}
\usepackage{filecontents}
\usepackage{algorithm2e}
\usepackage{algorithmic}
\usepackage{siunitx}
\usepackage{tikz}
\usepackage{tikzscale}
\usepackage{tabularx} 

\usetikzlibrary{spy}
\usetikzlibrary{snakes,arrows,shapes}    
\usetikzlibrary{spy}
\usetikzlibrary{backgrounds}  
\usetikzlibrary{decorations}
\usetikzlibrary{positioning}
\usetikzlibrary{patterns}
\usetikzlibrary{calc} 

\usepackage{booktabs}   
\usepackage{makecell} 
\usepackage{colortbl}   

\usepackage{xspace}   
\usepackage{setspace}   
 
\usepackage{soul}
\usepackage{ulem}
\usepackage{currfile}
\usepackage{import}

\usepackage{authoraftertitle} 
\usepackage{authblk} 
\usepackage{cleveref}

\pgfplotsset{compat=1.8}

\newcommand{\findmax}[3]{
    \pgfplotstablesort[sort key={#2},sort cmp={float >}]{\sorted}{#1}%
    \pgfplotstablegetelem{0}{#2}\of{\sorted}%
    \let #3=\pgfplotsretval%
}

\pgfplotscreateplotcyclelist{myCycleListBlackAndWhite}
{%
  {black, mark=o},
  {black, mark=square},
  {black, mark=triangle},
  {black, mark=diamond}, 
  {black, mark=pentagon}, 
  {black, mark=*},
  {black, mark=square*},
  {black, mark=triangle*},
  {black, mark=diamond*}, 
  {black, mark=pentagon*}, 
}

\definecolor{darkgreen}{rgb}{0,0.4,0} 
\definecolor{darkbrown}{rgb}{0.5, 0.396, 0.09}

\definecolor{c1}{rgb}{0.0, 0.4196078431372549, 0.6431372549019608}
\definecolor{c2}{rgb}{1.0, 0.5019607843137255, 0.054901960784313725}
\definecolor{c3}{rgb}{0.6705882352941176, 0.6705882352941176,
0.6705882352941176} \definecolor{c}{rgb}{0.34901960784313724, 0.34901960784313724, 0.34901960784313724}
\definecolor{c4}{rgb}{0.37254901960784315, 0.6196078431372549,
0.8196078431372549} \definecolor{c}{rgb}{0.7843137254901961, 0.3215686274509804, 0.0}
\definecolor{c5}{rgb}{0.5372549019607843, 0.5372549019607843,
0.5372549019607843} \definecolor{c}{rgb}{0.6352941176470588, 0.7843137254901961, 0.9254901960784314}
\definecolor{c6}{rgb}{1.0, 0.7372549019607844, 0.4745098039215686}
\definecolor{c7}{rgb}{0.8117647058823529, 0.8117647058823529,
0.8117647058823529}

\pgfplotscreateplotcyclelist{myCycleListColor}
{%
  {blue, mark=o},
  {red, mark=square},
  {darkbrown, mark=triangle},
  {darkgreen, mark=diamond},
  {black, mark=pentagon},
  {blue, mark=*},
  {red, mark=square*},
  {darkbrown, mark=triangle*},
  {darkgreen, mark=diamond*},
  {black, mark=pentagon*},
}

\pgfplotsset{every axis/.append style= 
              {
                font=\small,
                mark size=2,
                line width = 0.1,
                legend style={font=\small, mark size=3, draw=none, fill=none},
                legend cell align=left,
                cycle list name=myCycleListColor,
              }
            }
            
\pgfplotstableset
{
  every odd row/.style={before row={\rowcolor[gray]{0.9}}},
  every head row/.style=
  {
    before row=
    {%
      \toprule
    },
    after row=\midrule
  },
  every last row/.style=
  {
    after row=\bottomrule
  }
}

\newif\ifdrawboundingbox

\usetikzlibrary{external} 
\tikzset{external/system call={pdflatex \tikzexternalcheckshellescape
-halt-on-error -interaction=batchmode -jobname "\image" "\texsource"}} 

\pgfkeys
{ 
  /pgf/number format/.cd,
  fixed,
  set thousands separator={\,},
  1000 sep in fractionals,
}

\newcolumntype{C}[1]{>{\centering\arraybackslash}m{#1}}
\newcolumntype{R}[1]{>{\raggedright\arraybackslash}m{#1}}
\newcolumntype{L}[1]{>{\raggedleft\arraybackslash}m{#1}}

\newcommand{\delete}[1]{\xspace}

\definecolor{Reviewer1}{rgb}{0.0, 0.0, 0.0}
\definecolor{Reviewer2}{rgb}{0.0, 0.0, 0.0}
\definecolor{Reviewer3}{rgb}{0.0, 0.0, 0.0}
\title{Image-based material characterization of complex microarchitectured additively manufactured structures}

\author[1]{N. Korshunova\thanks{\href{mailto:nina.korshunova@tum.de}{\texttt{nina.korshunova@tum.de}},
    Corresponding author}}
\author[1]{J. Jomo}
\author[3]{G. L\'{e}k\'{o}}
\author[2]{D. Reznik}
\author[3]{P. Bal\'{a}zs}
\author[1]{\newline S. Kollmannsberger}

 \affil[1]{Chair for Computation in Engineering,
 Technische Universit\"at M\"unchen,
  Arcisstr. 21, 80333 M\"unchen, Germany}
 \affil[2]{Research and Technology Center, Siemens AG, Corporate Technology, Siemensdamm 50, 13629 Berlin, Germany}
 \affil[3]{Department of Image Processing and Computer Graphics, University of Szeged, \'{A}rp\'{a}d t\'{e}r 2, H-6720 Szeged, Hungary}

\newcommand{\journal}{Computers \& Mathematics with Applications}
\newcommand{\publicationDate}{November 13, 2019}

\usepackage{fancyhdr}
\date{}
\fancypagestyle{plain}{%
  \fancyhf{}%
  \fancyfoot[L]{
  \vspace{-1.25cm} 
  \footnotesize{Preprint submitted to \textit{\journal{}}  \hfill
  \publicationDate
  \\
  }} }

\crefname{figure}{Fig.}{Fig.}
\crefname{equation}{Eq.}{Eq.}
\crefname{table}{Tab.}{Tab.}

\drawboundingboxtrue
\drawboundingboxfalse 

\newcommand*{\figref}[2][]{%
	\hyperref[{fig:#2}]{%
		Fig.~\ref*{fig:#2}%
		\ifx\\#1\\%
		\else
		\,#1%
		\fi
	}%
}
\definecolor{changes}{RGB}{0,0,0}
\definecolor{changez}{RGB}{0,0,0}

\begin{document}  

\normalem
\maketitle  
  
\vspace{-1.5cm} 
\hrule 
\section*{Abstract}
{
	Significant developments in the field of additive manufacturing (AM) allowed the fabrication of complex microarchitectured components with varying porosity across different scales. However, due to the high complexity of this process, the final parts can exhibit significant variations in the nominal geometry. Computer tomographic images of 3D printed components provide extensive information about these microstructural variations, such as process-induced porosity, surface roughness, and other undesired morphological discrepancies. Yet, techniques to incorporate these imperfect AM geometries into the numerical material characterization analysis are computationally demanding. In this contribution, an efficient image-to-material-characterization framework using the high-order parallel Finite Cell Method is proposed. In this way, a flexible non-geometry-conforming discretization facilitates mesh generation for very complex microstructures at hand and allows a direct analysis of the images stemming from CT-scans. Numerical examples including a comparison to the experiments illustrate the potential of the proposed framework in the field of additive manufacturing product simulation.
}
 \vspace{.2cm} 
\vspace{0.25cm}\\
\noindent \textit{Keywords:} additive manufacturing, image-based material characterization, CT, microarchitectured, geometrical defects, finite cell method, deep learning, segmentation, Convolutional Neural Networks
 
\vspace{0.35cm}
\hrule 
\vspace{0.15cm}
\captionsetup[figure]{labelfont={bf},name={Fig.},labelsep=colon}
\captionsetup[table]{labelfont={bf},name={Tab.},labelsep=colon}
\tableofcontents
\vspace{0.5cm}
\hrule 
	\section{Introduction}
\label{sec:introduction}
{ 
	The production of complex customized structures using 3D printing has drawn significant interest both in industry and research due to its high flexibility and wide range of applications. A large spectrum of scientific developments has led to a better understanding of the underlying physical processes and has contributed to the improvement of additive manufacturing (AM) (e.g. \cite{Kollmannsberger2019}, \cite{Ngo2018}, \cite{Scott2012}). However, a complete description of the AM processes and products remains a challenge. 
	
	One particularly interesting research area is the material characterization of the final products, whose mechanical properties highly depend on the AM process parameters. Slight changes in these parameters may lead, for example, to surface roughness (e.g. \cite{Delgado2012}, \cite{Fox2016}, \cite{Spierings2011}), undesirable porosity or other microstructural defects (e.g. \cite{Gong2014}, \cite{Kasperovich2016}, \cite{Zhang2017}). Furthermore, it is more challenging to control these parameters for AM porous structures to achieve the predefined nominal microstructure (e.g. \cite{Maconachie2019}, \cite{RahmanRashid2017}, \cite{Tsopanos2010}). All these process-induced effects might reduce the mechanical properties of the final parts compared to conventionally manufactured ones. Therefore, the reliable evaluation of the mechanical behavior of AM components is one of the crucial tasks for defining their reliability and applicability. Yet, experimental mechanical testing in AM is limited by the high cost of printing and the need for a statistically significant number of repetitions using the same specimens. Thus, the robust estimation of the mechanical characteristics of the final AM products through computational methods is an important area of research.
	
	There is a considerable amount of literature on the determination of the linear and non-linear mechanical characteristics of fully dense AM materials, such as polymer or metals (see \cite{Herzog2016}, \cite{Ligon2017}, and literature cited therein). In particular, many studies have been carried out to investigate the influence of the printing direction (e.g. \cite{Simonelli2014}), surface roughness (e.g. \cite{Wycisk2014}) and microdefects on the final mechanical behavior of these materials (e.g. \cite{Gong2015}).
	
	Compared to fully dense AM materials, the macroscopic behavior of additively microarchitectured components, e.g. periodic lattices, highly depends on the achieved geometric accuracy of their microstructure (e.g. \cite{Dallago2019a}, \cite{Dong2019}, \cite{Liu2017}, \cite{Pasini2019}). Hence, for the numerical material characterization of such parts, a multi-scale mechanical analysis is usually carried out. This introduces another level of complexity to the numerical workflow. A straightforward approach to evaluate the macroscopic (effective) mechanical characteristics of these structures is the direct numerical simulation (DNS) of a virtual experiment. Another approach is numerical homogenization, which is advantageous due to its low computational cost. While the field of numerical material characterization has been widely studied, techniques to incorporate additive manufacturing defects of microarchitectured components have not yet been addressed in detail. One recently proposed approach is to perform numerical analysis on a statistically representative geometrical model determined from a computed tomography (CT) of final components \citep{Liu2017}.
	
	The present paper aims to propose an alternative direct image-based computational framework for the linear material characterization of imperfect microarchitectured AM products. We demonstrate the possibility of an efficient CT-based analysis of defective AM geometries using the high-order Finite Cell Method (FCM) \citep{Duster2017}. This numerical tool is used for two main ways to perform material characterization: direct numerical tensile testing and numerical homogenization. The CT-scans provide a comprehensive description of process-induced geometrical defects and can be directly integrated into the proposed approach without the need for a tedious mesh generation. Thus, it allows a flexible image-to-material-characterization workflow applicable to both periodic and random microarchitectured structures. 
	
	The structure of the paper is as follows. \Cref{sec::FCM} gives a brief overview of the numerical tools needed to establish an image-based material characterization workflow. First, the main ideas of the Finite Cell Method with a short description of the used voxel-based pre-integration technique are summarized. Next, a parallel computational framework is presented, which is essential for numerical testing of a complete specimen. Finally, the deep learning segmentation algorithm used to tackle the problems related to the metal CT-Scans is briefly discussed. \Cref{sec::numericalHomogenization} summarizes the concept of the mean-field numerical homogenization technique, which allows to reduce computational costs of the material characterization. Here, the advantages of the embedded approach are highlighted. In the fourth section, two simple case studies are shown to emphasize the effects of the boundary conditions on the resulting homogenized elastic properties arising from the numerical homogenization theory. Finally, in~\cref{sec::AMExamples}, the proposed framework is validated on two additively manufactured microarchitectured samples of Inconel\textregistered 718. 

}
	\section{The high-order Finite Cell Method for image-based multi-scale problems}
\label{sec::FCM}
\subsection{Concept of the Finite Cell Method}
\label{subsec::FCM}
{
	For the discretization of geometrically and topologically complex micro-structures stemming from computer tomographic images, it is advantageous to employ embedded techniques to avoid labor-intensive meshing procedures. In the scope of this work, the Finite Cell Method (FCM) is used. FCM is an embedded domain technique of high order and was presented in~ \cite{Duster2008, Duster2017, Parvizian2007}. In the following, only the main ideas of the FCM and its application to the direct numerical testing and homogenization in the scope of linear elasticity are recapitulated. 
	
	The weak form of the linear elastic governing equations in the absence of body forces may be written as follows:
	\begin{equation}
	\text{Find}  \,\,\,  \bm{u} \in H^1_{\hat{u}}(\Omega)  \,\,\,  \text{such that}  \,\,\, 
	\mathcal{B}(\bm{v,u}) = \mathcal{F}(\bm{v}) \,\,\, \forall \bm{v} \in H^1_0(\Omega)
	\label{eq::weakFormulationBVP}
	\end{equation}
	\hspace{-2mm}\begin{tabular}{p{.08\linewidth}p{.85\linewidth}}
		where
		&$\mathcal{B}(\bm{v,u}) = \int\limits_{\Omega} \bm{\varepsilon(v)}:\bm{C(x)}:\bm{\varepsilon(u)} \, d\Omega$\\
		&$\mathcal{F}(\bm{v}) = \int\limits_{d\Omega} \bm{v} \cdot \bm{t}\,d\Gamma_N$
	\end{tabular}
	\vspace*{1mm}
	
	The function spaces are defined as follows:
	\begin{equation}
	\begin{aligned}[c]
	&H^1_{\hat{u}}(\Omega) = \{ v_i \in H^1(\Omega): v_i = \hat{u}_i \,\,\, \forall \bm{x} \in \Gamma_D \}\\
	&H^1_{0}(\Omega) = \{ v_i \in H^1(\Omega): v_i = 0 \,\,\, \forall \bm{x} \in \Gamma_D \}
	\end{aligned}
	\end{equation}
	where $H^1$ denotes the Sobolev space of degree one, $\hat{u}$ is the prescribed displacement on the Dirichlet boundary $\Gamma_D$ , and $t$ is the prescribed surface traction on the Neumann boundary $\Gamma_N$.
	
	The basic idea of FCM is to embed the physical domain $\Omega$ into a regular-shaped domain $\Omega_e$ with vanishing stiffness (see~\cref{fig::FCMConcept}). To this end, the bilinear form in~\cref{eq::weakFormulationBVP} is modified as follows:
	\begin{equation}
	\mathcal{B}_e^\alpha(\bm{v,u}) = \int\limits_{\Omega_e} \bm{\varepsilon(v)}:\alpha(\bm{x})  \bm{C}(\bm{x}):\bm{\varepsilon(u)} \, d\Omega_e
	\label{eq::FCMBVP}
	\end{equation}  
	\noindent where $\alpha(\bm{x}) = \begin{cases}
	\phantom{\approx}1.0 \,\, \forall \,\bm{x} \in \Omega \\
	\approx 0.0 \,\, \forall \,\bm{x} \in \Omega_e\backslash \Omega
	\end{cases}$
	
	\begin{figure}[H]
		\centering
		\def\svgwidth{0.9\textwidth}
		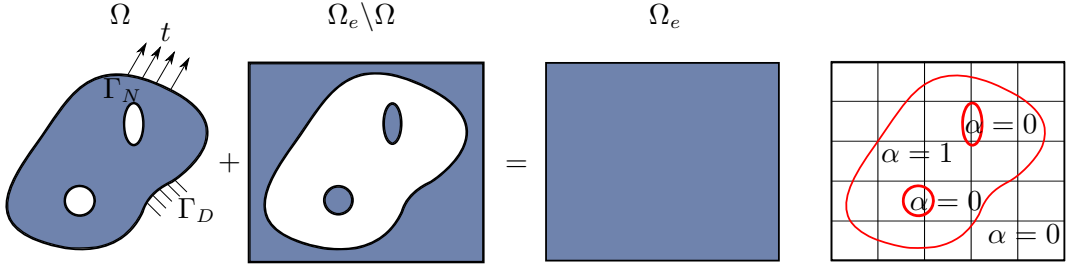
		\caption{Concept of the Finite Cell Method (adapted from \citep{Duster2017}).}    
		\label{fig::FCMConcept}
	\end{figure}

	The indicator function $\alpha(\bm{x})$ ensures that~\cref{eq::FCMBVP} is equivalent to ~\cref{eq::weakFormulationBVP} in terms of energy up to a modeling error proportional to $\sqrt{\alpha}$. The alpha is chosen  such that a compromise between the modeling error and the additional numerical error induced by a large condition number of the resulting system of equations is found. From a mechanical point of view, the value of $\alpha(\bm{x}) \,\, \forall \,\bm{x} \in \Omega_e\backslash \Omega$ corresponds to the addition of a material with vanishing stiffness in the void domain $\Omega_e\backslash \Omega$. 
	
	While the Finite Cell Method has been applied to a large variety of geometrical models (see e.g.~\cite{Rank2012, Wassermann2017, Wassermann2019}), the paper at hand focuses on an efficient voxel-based pre-integration technique introduced by Yang~\cite{Yang2012, Yang2012a}. This approach is a powerful tool to analyze image-based geometries for additive manufacturing applications.  
}

\subsection{An efficient voxel-based pre-integration technique}
\label{subsec::PreIntegration}
{    
	For numerical domains stemming from CT-scans, the computational efficiency of the FCM can be optimally exploited using a pre-integration technique on a voxel level \textcolor{Reviewer1}{introduced in~\cite{Yang2012a}. Only the main conceptual steps are addressed.}
	Consider a 3D voxel-based domain \textcolor{Reviewer1}{of size $l_x \times l_y \times l_z$ mm}. A slice of this domain is depicted in~\cref{fig::VoxelBasedPreIntegrationConcept} as an example. The numerical domain is discretized with $n_x \times n_y \times n_z$ finite cells. Every cell contains $v_x \times v_y \times v_z$ voxels. \textcolor{Reviewer1}{The voxel dimensions are $s_x \times s_y \times s_z$ mm.}   
	\begin{figure}[H]
		\centering
		\def\svgwidth{0.8\textwidth}
\begingroup%
  \makeatletter%
  \providecommand\color[2][]{%
    \errmessage{(Inkscape) Color is used for the text in Inkscape, but the package 'color.sty' is not loaded}%
    \renewcommand\color[2][]{}%
  }%
  \providecommand\transparent[1]{%
    \errmessage{(Inkscape) Transparency is used (non-zero) for the text in Inkscape, but the package 'transparent.sty' is not loaded}%
    \renewcommand\transparent[1]{}%
  }%
  \providecommand\rotatebox[2]{#2}%
  \newcommand*\fsize{\dimexpr\f@size pt\relax}%
  \newcommand*\lineheight[1]{\fontsize{\fsize}{#1\fsize}\selectfont}%
  \ifx\svgwidth\undefined%
    \setlength{\unitlength}{1847.03895869bp}%
    \ifx\svgscale\undefined%
      \relax%
    \else%
      \setlength{\unitlength}{\unitlength * \real{\svgscale}}%
    \fi%
  \else%
    \setlength{\unitlength}{\svgwidth}%
  \fi%
  \global\let\svgwidth\undefined%
  \global\let\svgscale\undefined%
  \makeatother%
  \begin{picture}(1,0.51897249)%
    \lineheight{1}%
    \setlength\tabcolsep{0pt}%
    \put(0,0){\includegraphics[width=\unitlength,page=1]{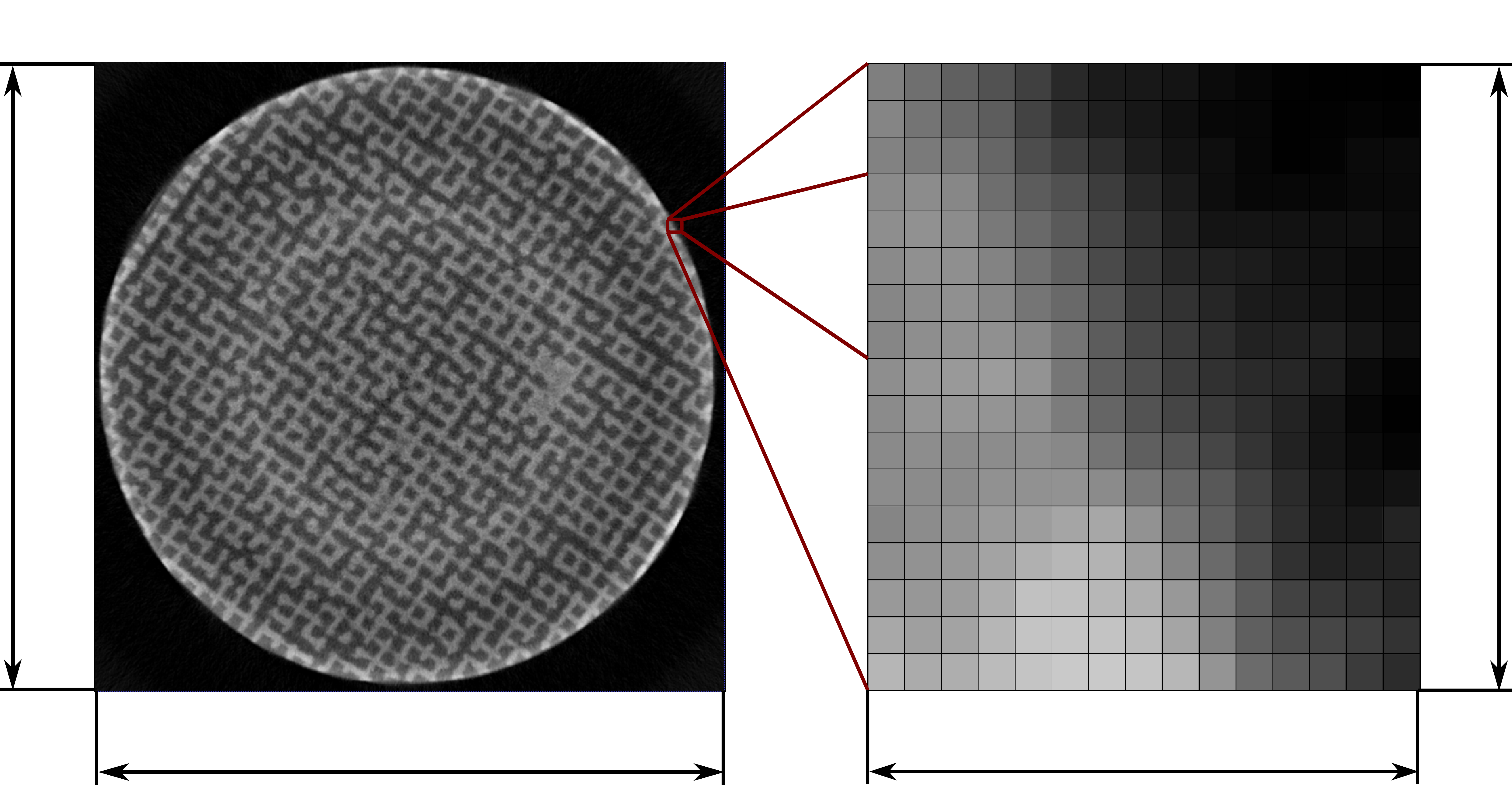}}%
    \put(0.18,0.02){\color[rgb]{0,0,0}\makebox(0,0)[lt]{\lineheight{1.25}\smash{\begin{tabular}[t]{l}$l_x=n_x\cdot v_x\cdot s_x$\end{tabular}}}}%
    \put(0.72,0.02){\color[rgb]{0,0,0}\makebox(0,0)[lt]{\lineheight{1.25}\smash{\begin{tabular}[t]{l}$v_x\cdot s_x$\end{tabular}}}}%
    \put(0.02,0.36){\color[rgb]{0,0,0}\rotatebox{-90}{\makebox(0,0)[lt]{\lineheight{1.25}\smash{\begin{tabular}[t]{l}$l_y=n_y\cdot v_y\cdot s_y$\end{tabular}}}}}%
    \put(0.97889998,0.23){\color[rgb]{0,0,0}\rotatebox{90}{\makebox(0,0)[lt]{\lineheight{1.25}\smash{\begin{tabular}[t]{l}$v_y\cdot s_y$\end{tabular}}}}}%
    \put(0.69,0.5){\color[rgb]{0,0,0}\makebox(0,0)[lt]{\lineheight{1.25}\smash{\begin{tabular}[t]{l}Finite Cell\end{tabular}}}}%
    \put(0.12,0.5){\color[rgb]{0,0,0}\makebox(0,0)[lt]{\lineheight{1.25}\smash{\begin{tabular}[t]{l}2D Slice of a CT-Scan\end{tabular}}}}%
  \end{picture}%
\endgroup%

		\caption{2D Slice of a CT-Scan with an example of a finite cell.}    
		\label{fig::VoxelBasedPreIntegrationConcept}
	\end{figure}
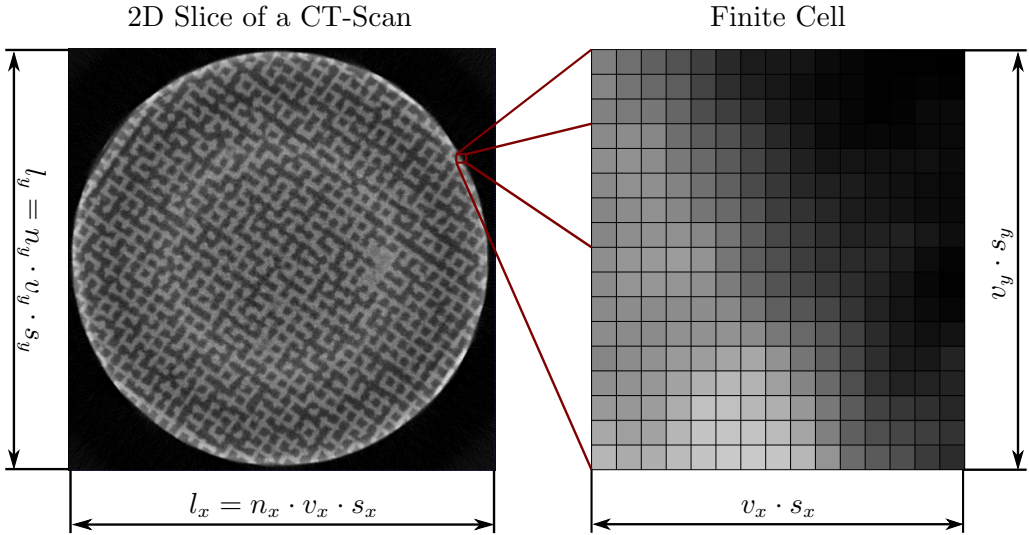
	 \textcolor{Reviewer1}{The Hounsfield units are piecewise constant in a voxel. Hence, the integrand in~\cref{eq::FCMBVP} is discontinuous at the boundary of the voxels within every finite cell. Therefore, the integration is carried out piecewise i.e. in a composed fashion on each voxel separately. The composed integration of the cell stiffness matrix after discretizing~\cref{eq::FCMBVP} then reads}:
	
	\begin{equation}
	\bm{K}_{c} = \int\limits_{\Omega_c} \bm{B^{T}}\left(\alpha(\bm{x}) \bm{C}(\bm{x})\right)\bm{B} \, d\Omega_c=\sum_{v_x}\sum_{v_y}\sum_{v_z}
	\int\limits_{\Omega_v} \bm{B^{T}}\left(\alpha(\bm{x}) \bm{C}(\bm{x})\right)\bm{B} \, d\Omega_v
	\label{eq::IntegrationFCMDecomposed}
	\end{equation}  
	where $\Omega_c$ is the finite cell domain and $\Omega_v$ is the domain of one voxel.

	\textcolor{Reviewer1}{Transferring~\cref{eq::IntegrationFCMDecomposed} to a local coordinate system of one finite cell, the integral can be written as follows}:
	\begin{equation}
	\bm{K}_{c} = \int\limits_{-1}^{1} \int\limits_{-1}^{1} \int\limits_{-1}^{1} \bm{B^{T}}\left(\alpha(\bm{x}) \bm{C}(\bm{x})\right)\bm{B} det \bm{J}\, d\xi d\eta d\zeta=\sum_{v_x}\sum_{v_y}\sum_{v_z} \left(\int\limits_{t_\xi}\int\limits_{t_\eta} \int\limits_{t_\zeta} \bm{B^{T}}\left(\alpha(\bm{x}) \bm{C}(\bm{x})\right)\bm{B} det \bm{J}\, d\xi d\eta d\zeta \right)
	\label{eq::IntegrationFCMDecomposedLocal}
	\end{equation}  
	where $t_\xi$,$t_\eta$ and $t_\zeta$ indicate the local coordinate limits of integration for every voxel within a finite cell. They depend on the number of considered voxels per dimension $v_x \times v_y \times v_z$.
	
	\textcolor{Reviewer1}{In the field of bio-mechanics, many methods have been proposed (see e.g~\cite{Yosibash2007}) to find an appropriate mapping between the Hounsfield units and the material coefficients. However, the paper at hand focuses on metal specimens produced by additive manufacturing for which such a mapping has not yet been developed. Therefore, constant homogeneous material properties are assumed. Thus, $\bm{C}(\bm{x})$ does not depend on spatial coordinates but is considered to be the constant tensor $\bm{C}$. The indicator function $\alpha(\bm{x})$ distinguishes between metal for which $\alpha(\bm{x}) = 1$ and void for which $\alpha(\bm{x}) \approx 0$. One way to obtain $\alpha(\bm{x})$ is to apply a threshold on the Hounsfield units of the CT scan. All voxels above the $HU_{thresh}$ are then marked to give $\alpha(\bm{x}) = 1$ and those below $HU_{thresh}$  are assigned to deliver zero. However, in the case of metal CT scans, this rather simple technique of applying a global threshold delivers bad results due to numerous kinds of imaging artifacts. Therefore, we apply the segmentation technique based on Convolutional Neural Networks described in more detail in~\cref{subsec::segmentation}}.
	
	Then, exploiting the structure of the integral indicated in~\cref{eq::IntegrationFCMDecomposedLocal}, the voxel stiffness matrix contributions can be pre-computed using a standard $(p+1)$ Gaussian quadrature rule. These contributions are then multiplied with the indicator function $\alpha(\bm{x})$ during the global stiffness matrix assembly depending on the location of the considered voxel.
	
	This pre-integration technique allows for fast integration of the global stiffness matrix. However, the storage of the dense $v_x \times v_y \times v_z$ matrices with the size of $d \times d$ increases the memory consumption, where $d$ indicates the number of degrees of freedom per finite cell (see~\cite{Yang2012a}). In the scope of the current work, the memory cost for storing these matrices is negligible compared to that of the rest of the simulation (e.g. the storage of the global stiffness matrix).
}

\subsection{A parallel computational framework for the Finite Cell Method}
{    
	As a high resolution of the microarchitectured geometry is necessary to accurately simulate the mechanical response of additively manufactured structures \textcolor{Reviewer1}{via direct numerical testing}, the size of the considered numerical problems rapidly increases and becomes prohibitively large so that they can no longer be computed on a single machine. To tackle this large-scale numerical simulation, a hybrid parallelization of the high-order FCM is used. This parallelization strategy is based on a modified version of the approach presented in \cite{Jomo2016, Jomo2019}. 
	
	
		\textcolor{Reviewer1}{The framework proposed in~\cite{Jomo2016} and employed in~\cite{Jomo2019} is replicating the entire mesh data structure on all participating MPI processes i.e. every MPI process knows the entire finite cell mesh in all detail and only selects a subset of these elements for further processing. This method was employed in~\cite{Jomo2016,Jomo2019} due to its simplicity because it automatically delivers consistent management of all degrees of freedom of the discretization. However, it not only requires a large amount of memory in total but the size of the largest computable structure is limited by the amount of memory one single MPI process can allocate. This memory limitation does not allow for a computation of the examples discussed in~\cref{sec::AMExamples}.} \textcolor{Reviewer1}{The limitations of the approach presented in~\cite{Jomo2016,Jomo2019} is lifted by the distributed mesh handling strategy. This strategy is based on a fully-parallel adaptive Cartesian mesh and is similar to the parallel mesh generation routines in packages such as P4est\cite{Burstedde2011}.}
\textcolor{Reviewer3}{The highly optimized parallel linear algebra capabilities of \texttt{Trilinos} for parallel matrix-vector multiplications and matrix storage are extensively used in this computational framework. Furthermore, the parallel Conjugate Gradient solvers from Trilinos are extended with the custom additive Schwarz based preconditioner, which treats the conditioning problems associated with the cut elements in FCM~\cite{Jomo2019,dePrenter2017a}.}

}
\subsection{CT-Images segmentation using Convolutional Neural Networks}
\label{subsec::segmentation}
{
	 In the scope of this work, the geometrical model for numerical analysis is obtained via computer tomography. The CT-images provide an extensive description of the process-induced morphological differences, thus, making the material characterization more accurate. However, it is well known, that different artifacts can be encountered in the CT-images \cite{Boas2012}. Ring artifacts, noise, beam hardening, etc. can obscure geometrical borders and make numerical analysis difficult. Furthermore, the presence of metal objects in the CT-scan causes severe metal artifacts. They arise due to the high attenuating material properties of the metal itself and the metal edges \cite{Boas2011, Man1998}. The question of artifacts reduction for a better reflection of the real objects is a separate area of research. Many different methods, e.g. projection completion strategies, multidimensional adaptive filtering, Metal Deletion technique, were proposed (\cite{Boas2012, Kalender1987, Prell2010}, and the literature cited therein). As the main focus of this contribution is the analysis of AM metal parts, the removal of the above-mentioned artifacts from received CT-images is essential for a reliable material characterization workflow. \textcolor{Reviewer1}{To this end, a simple threshold as described in~\cref{subsec::PreIntegration} does not suffice. Instead, we propose the deep learning segmentation presented in this section to obtain an accurate geometrical model for numerical analysis by providing a mapping between the Hounsfield units and the indicator function $\alpha(\bm{x})$.}

     Nowadays, Artificial Neural Networks (ANNs) are the best-known tools of artificial intelligence and machine learning. ANNs are widely used in a variety of ways when a task is needed to be learned. Convolutional Neural Networks (CNNs) \cite{Krizhevsky2017} are different from the structure of other neural networks since CNNs include mainly image processing functions and can also handle different types of input (e.g. image, video, voice). A typical use case of CNNs is to provide image data (on a pixel level) to the network as input and, based on this information, to perform classification. This is especially suitable for this work, as only the point membership classification is necessary to integrate the images into the proposed embedded workflow. This information is, then, used to determine the indicator function $\alpha(x)$ in \cref{eq::FCMBVP}. 
    
    For the segmentation of CT-image slices, in this case, a U-Net \cite{Ronneberger2015} based deep convolutional network (consisting of 6 blocks containing convolutional, max-pooling, dropout and merge layers) is optimized. The architecture is implemented in Keras \cite{chollet2015keras}. The Rectified Linear Unit (ReLU) activations are utilized, except for the last layer which has a sigmoid function. For optimization, an adaptive moment estimation (Adam) \cite{Kingma2014} with a binary cross-entropy loss function is employed. Depending on the task, 3-4 original slices (meaning less than 1 percent of them) have been manually segmented to get golden standards. The training data is then formed by these few original CT-slices together with their corresponding golden standard binary masks. The test dataset consists of the non-segmented slices. 
    
    To increase the number of training images, the input data is cut into pixel patches of size $n\times n$, where $n$ is chosen from the interval $[50,100]$ such that the remainder of the integer division $image\_width/n$ is the minimal possible. After the classification, the CT slices are reassembled from the patches. Since some false negatives and false positives may appear on the outer parts of the patches, they are padded on all sides to have a result of $100\times100$ pixel size. The specimen may contain some powder in it, which has a very similar intensity to the foreground pixels, so it might cause wrong classification results on a pixel level. To overcome this problem, some patches – in which powder is present – are shown twice to the neural network.
    
    The output of the network is a probability distribution, where the outputted values are expected to be very close to 0 or 1, so they can be interpreted as binary pixel values. As a post-processing step, global thresholding is applied with different threshold values and the qualitatively best has been chosen as the final binarized result. The inside-outside areas are indicated by these binary slices for further analysis.  \Cref{fig::CNNschematically} summarizes the main steps of the proposed segmentation approach. \textcolor{Reviewer1}{An
    overview of the five convolutional layers (depicted in~\cref{fig::CNNschematically}) can be seen in~\cref{fig::UNetScheme}. The type of operation and the number of convolutional kernels are noted above the boxes. The height, width and depth of the boxes are representative of the convolutional layers' output shape.}
    
    \begin{figure}[h]
    	\includegraphics[width=\textwidth]{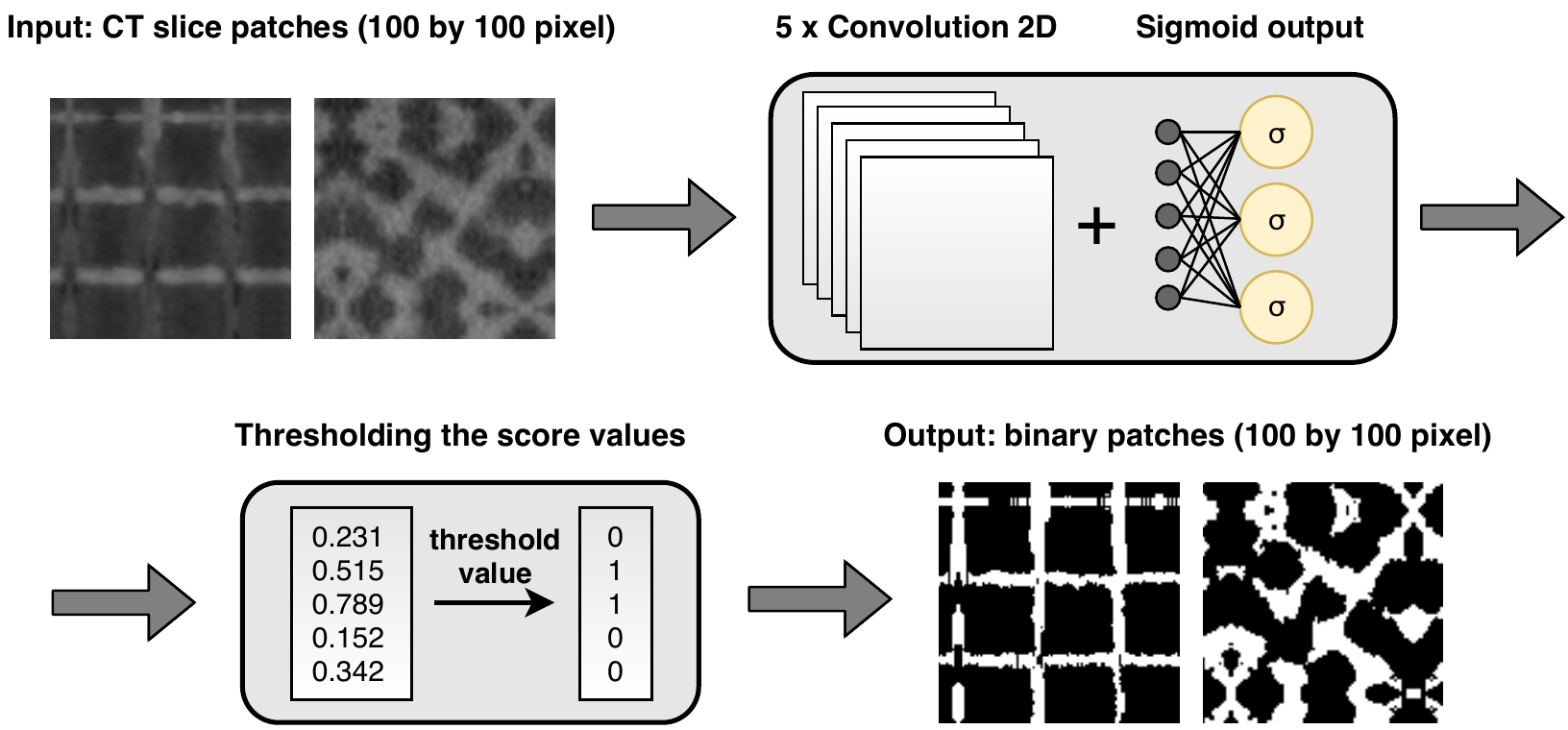}
    	\caption{Segmentation pipeline of CT-Scans of microarchitectured metal structures.}
    	\label{fig::CNNschematically}
    \end{figure}
    
    \begin{figure}[H]
    	\centering
    	\includegraphics[width=0.7\textwidth]{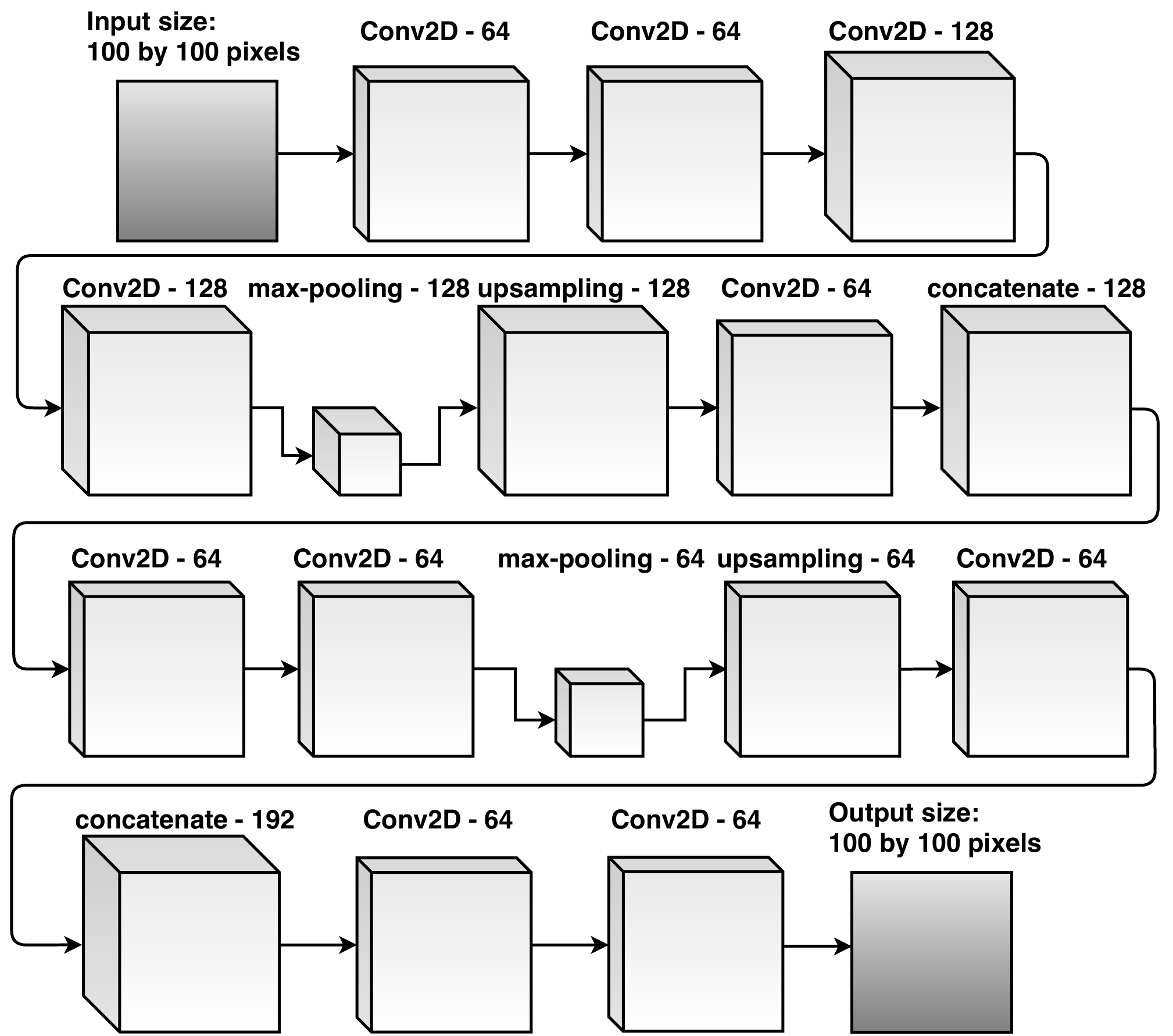}
    	\caption{An overview of the convolutional layers of the U-Net network.}
    	\label{fig::UNetScheme}
    \end{figure}
	
}
	\section{Numerical homogenization}
\label{sec::numericalHomogenization}
{    
	\textcolor{Reviewer1}{For a reliable numerical material characterization of microarchitectured materials generated by additive manufacturing it is of crucial importance to capture the "as-built" geometry and topology as accurately as possible. Then, one possibility is to use direct numerical tests on the full geometrical model. However, this requires large computational resources and the availability of large clusters of computers for every test, while delivering only few homogenized material constants. An alternative is offered by numerical homogenization. Homogenization is usually carried out on a much smaller representative volume element (RVE) which is computable on a personal computer. At the same time the complete homogenized elastic tensor can be computed without any assumptions on the macroscopic material law.} 
	
\textcolor{Reviewer1}{However, such an RVE is not easily defined in the microarchitectured foam-like structures generated by additive manufacturing under investigation in this paper. This is because there is a certain randomness in the produced microstructures themselves, which in turn means that no periodic unit cell can be defined for further computations.} 

\textcolor{Reviewer1}{To this end, we propose to sample the CT-Scan using many RVEs at different locations and to statistically characterize the obtained results. This strategy will be carried out in~\cref{sec::AMExamples} and compared to the computation of the entire structure performed by the methods described in the sections above. The main difficulty in this strategy is that a sampling of these somewhat random structures leads to voids randomly crossing boundaries of RVEs for which appropriate boundary conditions must be applied.}

\textcolor{Reviewer1}{We will demonstrate that the proposed numerical homogenization utilizing the Finite Cell Method (see~\cref{subsec::FCM}) in combination with an efficient pre-integration technique (see~\cref{subsec::PreIntegration}) and a CNN-based segmentation (see~\cref{subsec::segmentation}) for a generation of a reliable geometrical model provides a powerful tool for the material characterization of structures with a random microstructure. In the following, only the main aspects of homogenization are recapitulated with the focus on important aspects when using the embedded approach. In particular, we address the fact that the Finite Cell Method allows for an accurate and effortless incorporation of periodic boundary conditions in random RVEs and straightforward incorporation of the averaging relations in the numerical pipeline; a surprising but crucial feature for numerical homogenization.} 

\textcolor{Reviewer1}{To this end, we begin by discussing the transition from the microscopic to the macroscopic scale in~\cref{subsec::microToMacro} and focus on the accurate integration of the microscopic fields. \Cref{subsec::MacroToMicro} then reviews various boundary conditions to transfer the macroscopic quantities to the microscopic level. Special attention is paid to the advantages of the embedded approach in the application of the homogenization boundary conditions. The proposed numerical approach is not limited to the voxel-based geometries and can be applied to an arbitrary defined geometrical model, which does not necessarily need to comply with the periodic assumptions. The application of the numerical homogenization to periodic implicit unit cells will also be shown in the following sections.}
	\subsection{Micro-to-macro transition}
	\label{subsec::microToMacro}
	{
		For the transition of elastic quantities from the microscopic to the macroscopic scale, a representative volume element (RVE) is defined. The size of the RVE ($d$) should be chosen in such a way that it contains a sufficient number of heterogeneities ($l$) to represent an overall macroscopic behavior. At the same time, an RVE corresponds to a material point on the macroscopic scale ($L$). Therefore, the following equality should hold:
		
		\begin{equation}
		l<<d<<L
		\end{equation}
		
		Locally averaged microscopic fields over the RVE are, thus, required to be equivalent to an overall macroscopic value. 
		\textcolor{Reviewer1}{Often the RVE domain is composed of phases with piecewise constant material characteristics (e.g. inclusion, cavities, inhomogeneities etc.)}. Therefore, using the subdomain expansion the averaging theorems for stresses and strains can be written as follows\cite{Gross2017}:
		\begin{equation}
		\begin{aligned}[c]
		\bm{\sigma}^M &= \left< \bm{\sigma} \right>_\Omega =\sum_{\alpha=1}^{n} c_{\alpha} \left< \bm{\sigma} \right> _{\Omega_\alpha}= \frac{1}{\Omega} \int\displaylimits_{\Omega} \bm{\sigma}(\bm{x}) d\Omega\\%
		\bm{\varepsilon} ^M&= \left< \bm{\varepsilon} \right>_\Omega =\sum_{\alpha=1}^{n} c_{\alpha} \left< \bm{\varepsilon} \right> _{\Omega_\alpha}= \frac{1}{\Omega} \int\displaylimits_{\Omega} \bm{\varepsilon}(\bm{x}) d\Omega
		\end{aligned}
		\label{eq:averagingThreorem}
		\end{equation}
		\textcolor{Reviewer1}{
			\begin{tabbing}
				where \=$\alpha$ is a considered phase,\\
				\>$n$ is the total number of phases. i.e subdomains,\\
				\>$c_{\alpha}$ - volume fraction of a subdomain:\\
				\> $c_{\alpha} = \frac{V_{\alpha}}{V}$ and $\sum_{\alpha=1}^{n} c_{\alpha}=1$
			\end{tabbing}
		 }
		It is important to note that the volume $\Omega$ is the total microscopic volume occupied by the matrix $\Omega_{m}$ and the heterogeneities $\Omega_{h}$, i.e. $\Omega = \Omega_{m} \cup \Omega_{h}$. Such heterogeneities may be inclusions, voids, cracks, etc. (see~\cref{fig:volumesDefinitions}, where $m$ indicates matrix material and $h$ - heterogeneities).
		
		\begin{figure}[H]
			\centering
			\def\svgwidth{\linewidth}
			\Large{\scalebox{.6}{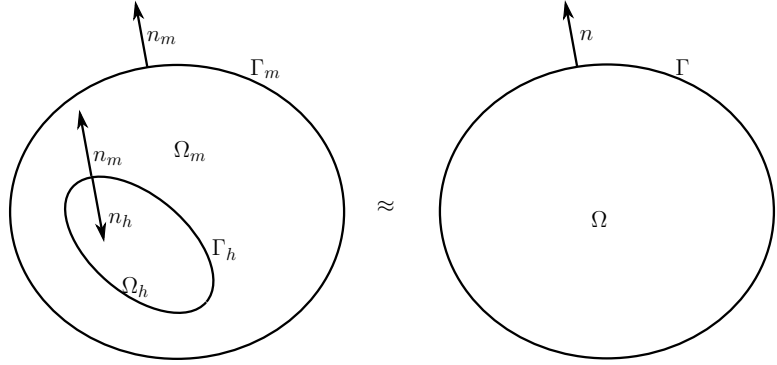}}
			\caption{Variables definition for an RVE as a macroscopic point}
			\label{fig:volumesDefinitions}
		\end{figure}
		
		If the volume $\Omega$ is compact, has a piecewise smooth boundary and the stress field $\bm{\sigma}$ is continuously differentiable within the domain $\Omega$, ~\cref{eq:averagingThreorem} for the average microscopic stresses can be formulated in terms of a surface integral:
		\begin{equation}
		\bm{\sigma^M} = \left< \bm{\sigma} \right>_\Omega = \frac{1}{\Omega} \int\displaylimits_{\Omega} \bm{\sigma}(\bm{x})d\Omega = \frac{1}{\Omega} \int\displaylimits_{d\Gamma} \bm{t} \otimes \bm{x} \,d\Gamma
		\label{eq:averagingThreoremSurfaceStress}
		\end{equation}
		where the following definition of traction forces is used:
		\begin{equation}
		\mathbf{t} = \bm{\sigma} \cdot \mathbf{n}
		\label{eq:directTractionEquation}
		\end{equation}
		
		The same argument holds for a continuously differentiable strain field $ \bm{\varepsilon} $ in the RVE $\Omega$:         
		\begin{equation}
		\bm{\varepsilon^M} = \left< \bm{\varepsilon} \right>_\Omega =\frac{1}{\Omega}\int\displaylimits_{\Omega}\bm{\varepsilon} (x)\,d\Omega = \frac{1}{\Omega}\int\displaylimits_{d\Gamma} (\mathbf {u} \otimes_s \mathbf {n} )\,d\Gamma
		\label{eq:averagingThreoremSurfaceStrains}
		\end{equation}
		
		The main assumption of continuously differentiable stress and strain fields does not necessarily hold for heterogeneous materials due to a material discontinuity over the boundary $\Gamma_h$ (see~\cref{fig:volumesDefinitions}). However, applying the sub-domain expansion as in~\cref{eq:averagingThreorem} and assuming equilibrium and displacement continuity along the surface  $\Gamma_h$ ($t_i^m \overset{!}{=} t_i^h$ and $u_i^h \overset{!}{=} u_i^m $), the general form of ~\cref{eq:averagingThreoremSurfaceStress,eq:averagingThreoremSurfaceStrains} remains valid.
		
		Furthermore, a continuous equilibrium is not necessarily established along  $d\Omega_h$ if the displacement-based Finite Element or Finite Cell approximations are applied to obtain the primary variables. While $C^0$ continuity across the element boundaries of the primary variables is guaranteed, only \textcolor{Reviewer1}{discontinuous stresses and strain fields} can be expected. Taking into consideration \cref{eq:averagingThreoremSurfaceStress,eq:averagingThreoremSurfaceStrains}, displacements and tractions at the boundary of evaluation need to be continuous. The direct computation of the tractions using~\cref{eq:directTractionEquation} in the boundary integral of \cref{eq:averagingThreoremSurfaceStress} involves a lower degree of the stress extrapolation than the order of the shape functions used for the discretization of the primary variables. This results in poor accuracy and a lower rate of convergence of the first derivatives of the primary variables. Moreover, the differential equilibrium is not satisfied at every point -- and will thus result in non-equilibrium of the tractions at the integration points over the boundary $\Gamma$. 
		
		This problem can be tackled by many approaches, for example using NURBS as basis functions to ensure higher continuity between the elements. However, as the nodal force equilibrium always holds for any finite element or finite cell mesh for standard approximations \cite{Bathe2007}, the following equation is used:
		
		\begin{equation}
		\left< \bm{\sigma} \right>_\Omega = \frac{1}{\Omega} \int\displaylimits_{d\Omega} \bm{t} \otimes \bm{x} \,d\Gamma = \frac{1}{\Omega} \sum\limits_{i=1}^{n_{nodes}} \bm{t} \otimes \bm{x}
		\end{equation}
		where
		
		\begin{equation}
		\bm{t}_{el} = \int\displaylimits_{\Omega_{el}} \mathbf{B}^{T}\bm{\sigma} \,d\Omega_{el} 
		\label{eq:correctTractionEvaluation}
		\end{equation}
		
		For numerical reasons, it is convenient to evaluate~\cref{eq:correctTractionEvaluation} after the computation of the Finite Cell solution as $\mathbf{K} \mathbf{u}$ and extract the corresponding nodal tractions during a postprocessing step.

		As most of the heterogeneities considered in the following chapters are voids, a material definition of a void is introduced. Following~\cite{Suquet1985} voids are assumed to be an infinitely soft heterogeneity with vanishing stiffness:        
		\begin{equation}
		\bm{C}_h=\bm{C}_v \approx \bm{0}
		\label{eq::VanishingStiffness}
		\end{equation}
		
		With this definition at hand, consider an RVE with a cavity completely enclosed in a volume (see~\cref{fig:volumesDefinitions}). The total volume of the RVE ($\Omega$) consists of the volume of the void ($\Omega_v=\Omega_h$) and the matrix ($\Omega_m$). The boundary of the void ($\Gamma_v=\Gamma_h$) does not intersect the boundary of the RVE ($\Gamma$), i.e. $\Gamma_v \cap \Gamma = \emptyset$. The total boundary of the volume of the matrix $\Gamma_m$ consists of two boundaries: an outer boundary which coincides with the boundary of the RVE $\Gamma$, and an interior boundary, which coincides with the boundary of the void. The normals are defined positive in case they point outwards of the volume they bound -- and as negative otherwise. 
		
		If the cavity is truly void, the strain field in the cavity is ambiguous. However, using the definition of the void from~\cref{eq::VanishingStiffness} and using the sub-domain expansion~\cref{eq:averagingThreorem}, the strain average over the total volume can be rewritten as follows:%
		\begin{equation}
		\bm{\varepsilon} ^M= \left< \bm{\varepsilon} \right>_\Omega = \frac{1}{\Omega} \left( \int\displaylimits_{\Omega_m} \bm{\varepsilon}(\bm{x}) d\Omega_m  + \int\displaylimits_{\Omega_v} \bm{\varepsilon}(\bm{x}) d\Omega_v \right)
		=c_m<\bm{\varepsilon}>_m +  \int\displaylimits_{\Gamma_v} (\mathbf {u} \otimes_s \mathbf {n} )\,d\Gamma_v=c_m<\bm{\varepsilon}>_m + \varepsilon_c 
		\label{eq:strainDefinitionInTheVoid}
		\end{equation}
		
		The strains within the void domain are called \textit{cavity strain} $\varepsilon_c$ (see Section 5 in \cite{nemat2013micromechanics}). The cavity strain is an average of additional strain fluctuations induced by the deformation of the boundary of the cavity compared to the deformation state of a purely homogeneous domain~\citep{Fritzen2013}. Thus, the cavity strain represents an equivalent eigenstrain as introduced by Eshelby (refer to~\cite{Gross2017} and \cite{nemat2013micromechanics}).
		
		The average macro-stress is equal to the weighted stress average in the matrix as the stress contribution from the infinitely soft heterogeneity is vanishing:
		
		\begin{equation}
		\begin{aligned}[c]
		\bm{\sigma} ^M= \left< \bm{\sigma} \right>_\Omega = \frac{1}{\Omega} \left( \int\displaylimits_{\Omega_m} \bm{\sigma}(\bm{x}) d\Omega_m  + \int\displaylimits_{\Omega_h} \bm{\sigma}(\bm{x}) d\Omega_h \right)=c_m<\bm{\sigma}>_m +  c_v\cancelto{\approx 0}{<\bm{\sigma}>_v}
		\end{aligned}
		\label{eq:stressDefinitionInTheVoid}
		\end{equation}
		
		Therefore, when cavities are completely inside an RVE and do not intersect with its boundaries, overall average strains and stresses are fully determined in terms of the fields residing at the outer boundary of the RVE $\Gamma_m$. 
		
		When an intersection between a boundary of a void and a boundary of an RVE is not empty, i.e. $\Gamma_v \cap \Gamma \neq \emptyset$, the averaging relations given in~\cref{eq:strainDefinitionInTheVoid,eq:stressDefinitionInTheVoid} still remain valid. Yet, its definitions in terms of boundary integrals become rather cumbersome as a split into $\Gamma_v \cap \Gamma$ and $ \Gamma /(\Gamma_v \cap \Gamma) $ needs to be considered.
		
		Nevertheless, the definition of the void as an infinitely soft medium provides a unique definition of the strain and stress fields in the required segment $\Gamma_v \cap \Gamma$.
		Using the Finite Cell Method described in \cref{sec::FCM} for this case is advantageous as it mimics the assumption of the voids being a material with vanishing stiffness (see~\cref{eq::VanishingStiffness}). The indicator function $\alpha_(\bm{x})$ in the Finite Cell Formulation can be interpreted as the inclusion of an infinitely soft material. A displacement field in the void domain is, thus, computed naturally. An extension of a displacement field to the void domain is required to be consistent with the formulation of the boundary value problem. Thus, the consistency of strain energy is preserved naturally in the Finite Cell Method \cite{Parvizian2007}. 
	}
	\subsection{Macro-to-micro transition}
	\label{subsec::MacroToMicro}
	{        
		Having defined a micro-to-macro coupling between the scales, a very brief review of the macro-to-micro transition in the homogenization theory is given. Let us assume linear elastic material and small strain on both the microscopic and the macroscopic scales. The general Hooke's law on a microscopic level is then expressed by:%
		\begin{equation}
		\bm{\sigma}(\mathbf{x}) =  \mathbf{C}(\mathbf{x}) : \bm{\varepsilon}(\mathbf{x})
		\label{eq:HookesLawMicroscopic}
		\end{equation}
		
		Given the definition of the effective material properties in~\cref{eq:averagingThreorem}, an effective elasticity tensor $C_{ijkl}^*$ is introduced relating the macroscopic fields:
		\begin{equation}
		\bm{\sigma}^M = \mathbf{C}^* : \bm{\varepsilon}^M
		\label{eq:HookesLawMacroscopic}
		\end{equation}
		
		To allow an interpretation of $\mathbf{C}^*$ as a material characteristic on the macroscopic level, the condition of strain energy equality must be satisfied \citep{Gross2017}. The average strain energy density in the RVE $\Omega$ must be equal to the strain energy density in the macroscopic point:
		
		\begin{equation}
		\frac{1}{2} \left(\bm{\sigma}^M : \bm{\varepsilon}^M \right) = \int\displaylimits_{\Omega}\frac{1}{2}\left(\bm{\sigma} : \bm{\varepsilon} \right)d\Omega
		\label{eq:energyEquivalence}
		\end{equation}
		
		\Cref{eq:energyEquivalence}, also known as Hill condition, can be rewritten in a more conventional form using the definitions in~\cref{eq:averagingThreorem}:
		\begin{equation}
		\left<\bm{\sigma}\right>_\Omega : \left<\bm{\varepsilon}\right>_\Omega = \left<\bm{\sigma} : \bm{\varepsilon} \right>_\Omega
		\label{eq:HillCondition}
		\end{equation}
		
		Let us assume that a split of total microscopic fields on an average and a fluctuating part exists:
		\begin{align}
		\bm{\sigma}( \mathbf{x}) &=\left<\bm{\sigma}\right>_\Omega +     \bm{\widetilde{\sigma}}(\mathbf{x})\\
		\bm{\varepsilon} (\mathbf{x}) & = \left<\bm{\varepsilon}\right>_\Omega +     \bm{\widetilde{\varepsilon}}(\mathbf{x})
		\end{align}		
		The Hill condition \cite{Hill1963} can be expressed in a different form using the field fluctuations:
		\begin{equation}
		\left<\bm{\widetilde{\sigma}}(\mathbf{x})\bm{\widetilde{\varepsilon}}(\mathbf{x}) \right>_\Omega = 0
		\label{eq:HillConditionFluctuations}
		\end{equation}
		
		\Cref{eq:HillConditionFluctuations} requires the stress fluctuations not to do any work on the strain fluctuations on average. The volume integral in~\cref{eq:HillConditionFluctuations} can be expressed in terms of boundary quantities:
		
		\begin{equation}
		\frac{1}{\Omega} \int\displaylimits_{d\Omega} \left(\mathbf{u} -  \left< \bm{\varepsilon}\right>_\Omega \mathbf{x} \right) \left(\bm{\sigma} -\left< \bm{\sigma}\right>_\Omega \right)\mathbf{n} \, d\Gamma = 0
		\label{eq:HillConditionAsBoundaryIntergral}
		\end{equation}
		where $\mathbf{u}$, $\bm{\sigma}$, and $\bm{\varepsilon}$ correspond to the microscopic displacement, the stress and the strain fields, respectively.
		
		\Cref{eq:HillConditionAsBoundaryIntergral} shows that the fluctuations of the micro-fields along the boundary of an RVE must be energetically equivalent to their averages. 
		
		To complete the formulation of the homogenization problem, appropriate boundary conditions need to be specified. They can be deduced from the Hill criteria written in the boundary form (see~\cref{eq:HillConditionAsBoundaryIntergral}). The following boundary conditions are widely used \citep{Pahr2008}:
		
		\begin{itemize}
			\item Kinematic uniform boundary conditions (KUBC, linear displacements)
			\begin{equation}
			\mathbf{u}|_{d\Omega} = \bm{\varepsilon}^M \mathbf{x}
			\label{eq:KUBC}
			\end{equation}
			\item Static uniform boundary conditions (SUBC, tractions)
			\begin{equation}
			\mathbf{t}|_{d\Omega} = \bm{\sigma}^M \mathbf{n}
			\label{eq:SUBC}
			\end{equation}
			\item Displacement periodic boundary conditions (PBC)
			\begin{align}
			\mathbf{u}(\mathbf{x^+})|_{d\Omega} - \mathbf{u}(\mathbf{x^-})|_{d\Omega}& = \varepsilon^M \Delta \mathbf{x}\nonumber\\
			\mathbf{t}(\mathbf{x^+}) &=-\mathbf{t} (\mathbf{x^-})
			\label{eq:PBC}
			\end{align}
			\item Mixed uniform boundary conditions (MUBC)
		\end{itemize}
		
		All of the mentioned boundary conditions satisfy the Hill condition a priori. The average stresses are only meaningful if self-equilibrating forces are applied through~\cref{eq:SUBC} (see \cite{nemat2013micromechanics}). However, in the case of the uniform kinematic condition, a requirement of displacements to be self-compatible is not necessary.
		
		The order relations of the apparent elasticity tensors driven by the different boundary conditions are well-established in the literature \cite{Hazanov1994}. An effective elasticity tensor is always bounded by the apparent tensors estimated by KUBC and SUBC:
		
		\begin{equation}
		\mathbf{C}_{SUBC} \leq \mathbf{C}^* \leq \mathbf{C}_{KUBC} 
		\label{eq:BCsAsBoundsToEffectiveTensor}
		\end{equation}
		
		Operator $\leq$ in \cref{eq:BCsAsBoundsToEffectiveTensor} for Voigt notation means that matrix quadratic forms:
		
		\begin{align}
		Q_1 &=  \bm{\sigma}^T \left(\mathbf{C}^* - \mathbf{C}_{SUBC}\right) \bm{\sigma}\\
		Q_2 &=  \bm{\sigma}^T \left(\mathbf{C}_{KUBC}^* - \mathbf{C}^*\right)  \bm{\sigma}
		\end{align} 
		should be positive semi-definite \cite{Hazanov1994}. The semi-positive definiteness of the quadratic forms can be determined either by the eigenvalues or the semi-positivity of the principal minors of the matrix $\left(\mathbf{C}^* - \mathbf{C}_{SUBC}\right)$.
			
		For periodic microstructures, the exact effective stiffness can be estimated with the application of PBC (see e.g. \citep{Pahr2003,Sanchez-Palencia1987}). The gap between $ \mathbf{C}_{KUBC}$ and $ \mathbf{C}_{SUBC}$ remains significant \cite{Suquet1985}. However, with the increase of the size of the RVE, all of the estimates must converge to one value. The order relation is then summarized as follows:
		
		\begin{equation}
		\mathbf{C}_{SUBC} \leq \mathbf{C}_{PBC}  \leq \mathbf{C}_{KUBC} 
		\label{eq:BCsAndPBCAsBoundsToEffectiveTensor}
		\end{equation}
		
		The periodic boundary conditions are widely used even if the periodicity requirement doesn't hold (see \cite{Nguyen2012a} and the literature therein). 
		
		The application of the PBC requires identical meshes on the polar RVE surfaces ($\bm{x^+}$ and $\bm{x}^-$). This becomes rather cumbersome and sometimes even impossible when the RVE has a non-periodic microstructure. One solution is to apply the PBC weakly, as described in~\cite{Nguyen2012a}. Yet, the Finite Cell Mesh consists of Cartesian grids with coupled nodes on all six surfaces of an RVE independently on its microstructure. This makes the application of the PBC natural and does not require any additional effort.
	}        
}

	\section{Numerical investigations}
\label{sec:numericalInvestigations}
{
	\newcommand{\graphDir}{./sections/examples/NumericalInvestigations/Pictures}
	In this section, the verification and investigation of the numerical homogenization technique are performed on two different examples. In~\cref{subsec::SphericalInclusion}, the results are verified for a spherical inclusion published in the literature. The second example is discussed in~\cref{subsec::CubicVoid}, where an RVE with a cubical void is considered. To this end, the three most common boundary conditions (KUBC, PBC, and SUBC) are applied to compare their effect on the macroscopic homogenized properties. 
	\subsection{Cubic unit cell with a single spherical particle}
	\label{subsec::SphericalInclusion}
	{
		In this section, in order to verify the current implementation, an example of a centered hard spherical inclusion inside a soft cubic unit cell is considered. The setup used for the numerical simulation is indicated in~\cref{fig::SphericalInclusion}. 
		\begin{figure}[H]
			\centering
			\captionsetup[subfigure]{oneside,margin={0cm,0cm},labelformat=empty,captionskip=40pt}
			\subfloat[(a) Setup]{\def\svgwidth{0.3\textwidth}
				\input{\graphDir/SphericleParticleSetup.pdf_tex}}
			\subfloat[(b) Displacement field ]{
				\includegraphics[width=0.4\textwidth, trim={6cm 2cm 10cm 2cm },clip]{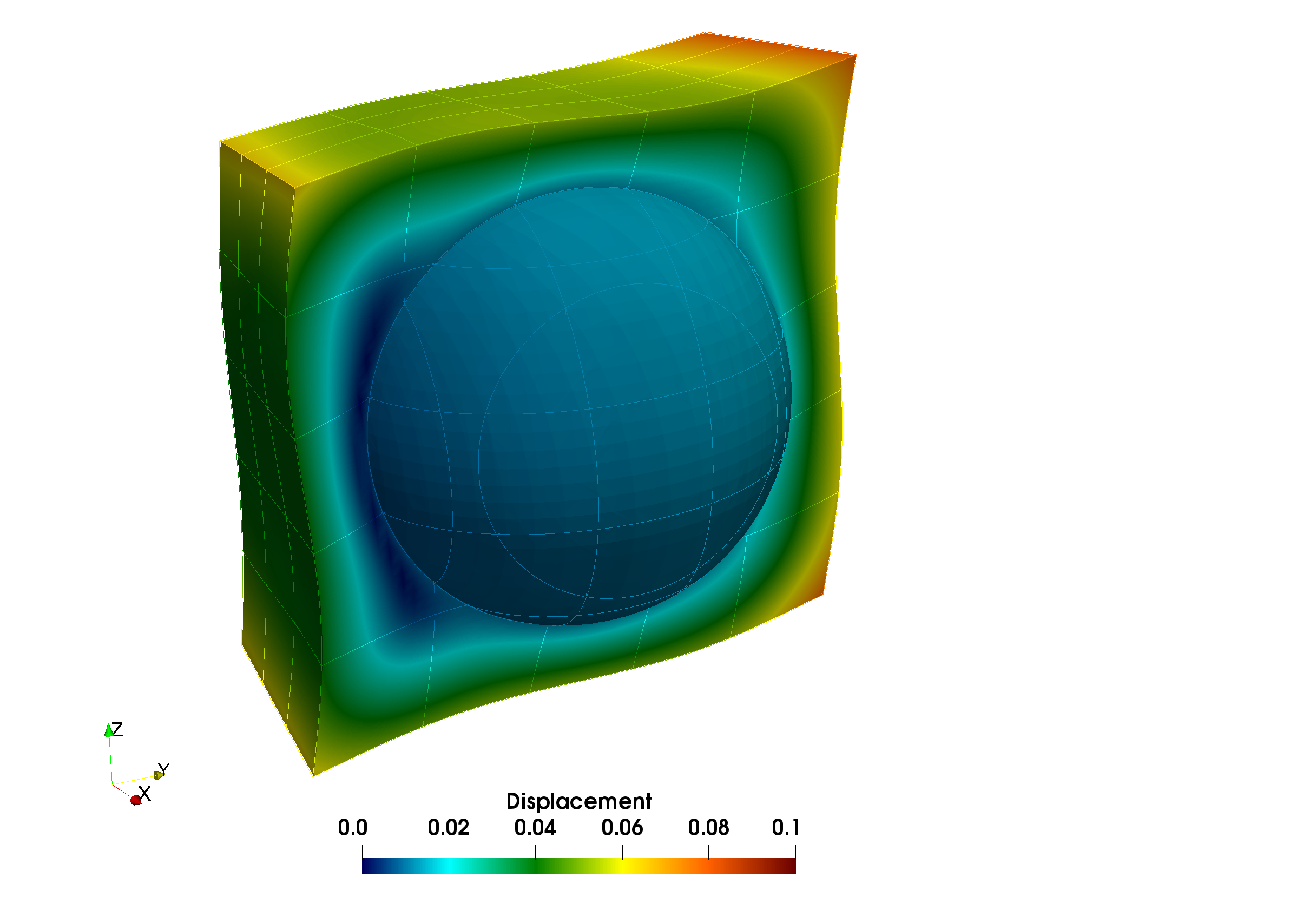}}        
			\hspace*{-1.5cm}
			\subfloat[(c) Von Mises stress distribution]{
				\includegraphics[width=0.4\textwidth, trim={6cm 2cm 10cm 2cm },clip]{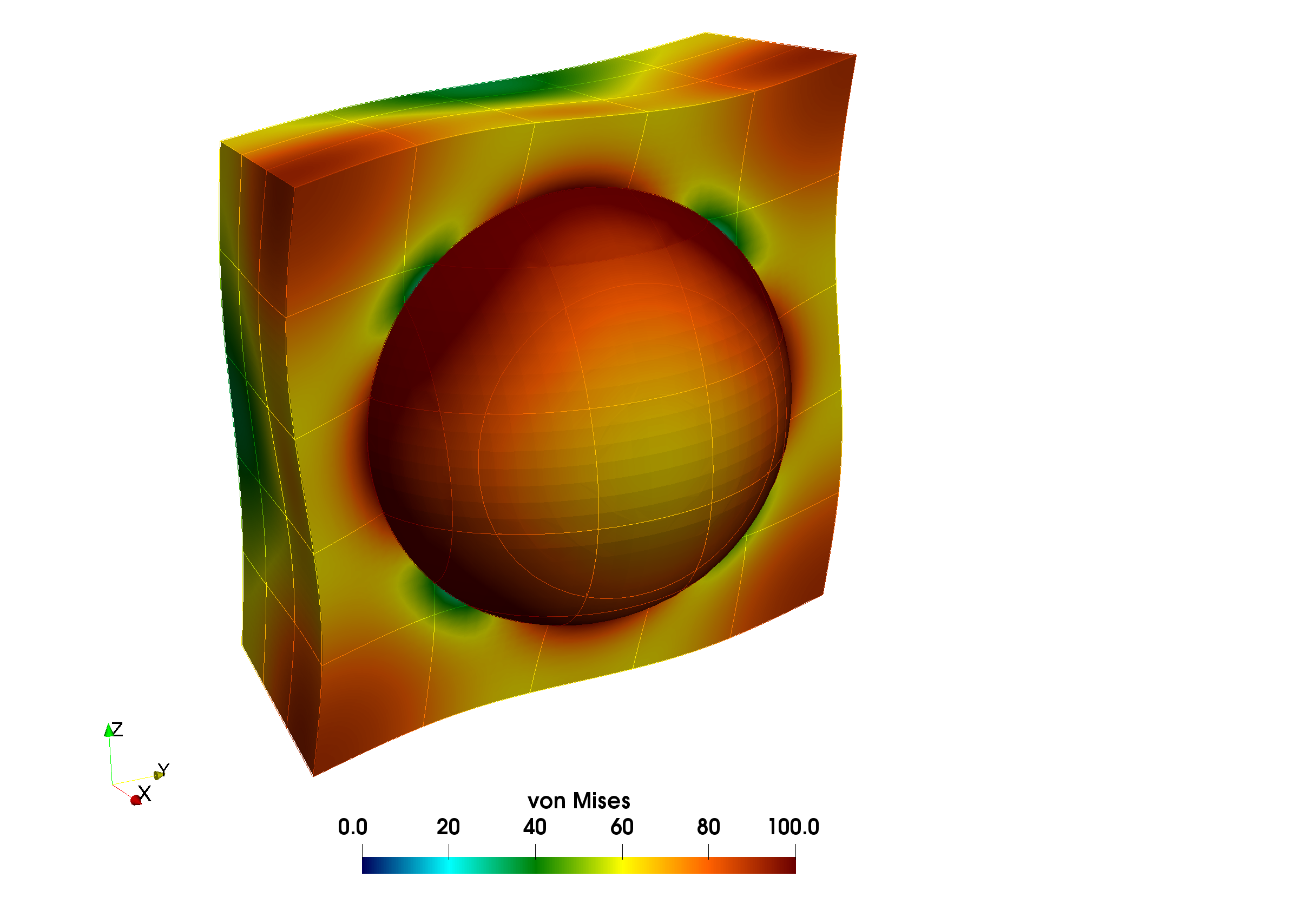}}
			\caption{Hard spherical inclusion in a soft matrix.}
			\label{fig::SphericalInclusion}
		\end{figure}        
		
		This periodic unit cell is computed using the high-order Finite Cell Method (\cref{sec::FCM}) in combination with the Smart-Octree integration described in~\cite{Kudela2016}. \textcolor{Reviewer3}{The convergences curves using $4\times4\times4$ and $5\times5\times5$ cells are presented in~\cref{fig::SphericalInclusionConv}. The polynomial degree is raised from $p=1$ to $p=7$.} Due to the presence of a material interface, two meshes are employed to discretize the problem. The interface condition between the two materials is enforced weakly with a penalty parameter of $\beta=10^8$, while the fictitious indicator function is kept as $\alpha = 10^{-9}$. The Periodic Boundary Condition (\cref{subsec::MacroToMicro}) is imposed conventionally, taking advantage of the symmetry of the Finite Cell meshes. 
		
		\begin{figure}[H]
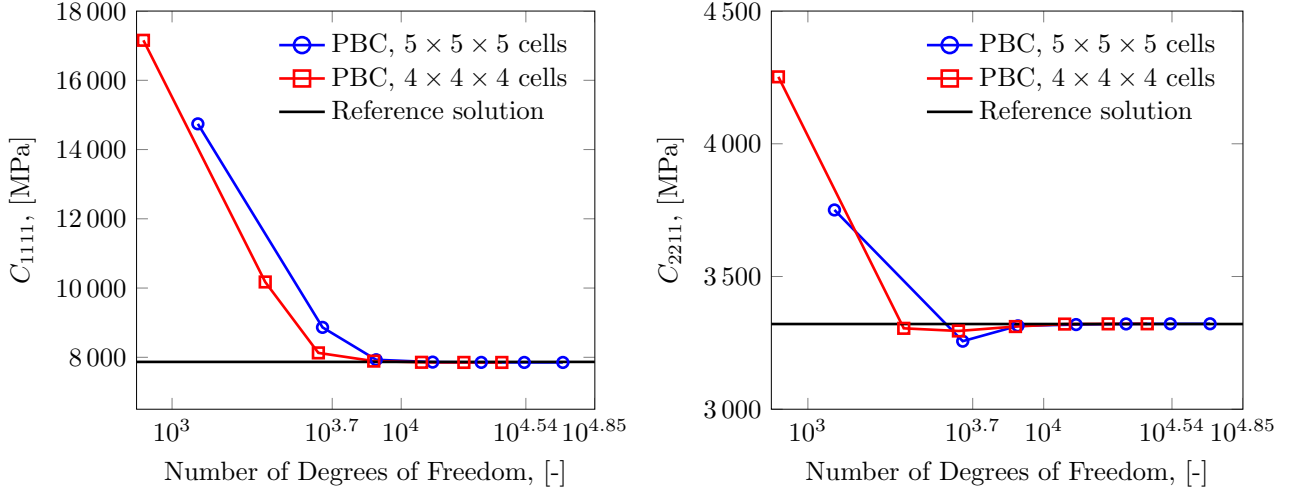

			\centering
			\captionsetup[subfigure]{oneside,margin={0cm,0cm},labelformat=empty}
			\subfloat[ ]{
				\includegraphics[width=0.5\textwidth,height=0.28\textheight]{\graphDir/TikZ/SphericalInclusion/graphs/pStudyC11.tikz}}
			\subfloat[ ]{	\includegraphics[width=0.5\textwidth,height=0.28\textheight]{\graphDir/TikZ/SphericalInclusion/graphs/pStudyC21.tikz}} \caption{Convergence studies of two entries into the elasticity tensor $C_{1111},C_{2211}$.}
			\label{fig::SphericalInclusionConv}
		\end{figure}     
	
		The reference solutions of this problem and the solution of the present work is summarized in \cref{tab::referenceSolutionSphericalInclusion} (refer to \cite{Gusev1997,Michel1999}).
		\begin{table}[H]
			\centering
			\begin{tabular}{| c | c | c | c | c | c | c | c | c | c|}
				\hline        
				& $c_p$ & $N$ & $C^*_{1111}$ &$C^*_{2211}$    &$C^*_{2222}$    &$C^*_{3333}$    &$C^*_{1212}$    &$C^*_{2323}$    &$C^*_{1313}$    \\    
				& $[-]$ & & $[GPa]$ &$[GPa]$    &$[GPa]$    &$[GPa]$    &$[GPa]$    &$[GPa]$    &$[GPa]$    \\\hline    
				\cite{Gusev1997} &  0.2678 & $\approx 3.4 \cdot 10^4$ &  8.069 & -&8.075 & 8.072 & 1.839 & 1.834 &1.835\\
				&& tet./sphere &&&&&&&\\
				\cite{Michel1999} & 0.2678 & $240^3$ pixels &  7.867 & 3.321&-&-&-&-&-\\
				Present work & 0.2678 & $50\,736$ DOFs &  7.851 & 3.321&7.851&7.851&1.780&1.780&1.780\\
				\hline    
			\end{tabular}
			\caption{Two-phase composite: homogenized elastic constants (\cite{Gusev1997}, \cite{Michel1999})}
			\label{tab::referenceSolutionSphericalInclusion}
		\end{table}

		\textcolor{Reviewer3}{\Cref{tab::referenceSolutionSphericalInclusion} and \cref{fig::SphericalInclusionConv} show that the Finite Cell Method in combination with the Smart-Octree integration benefits from an exact resolution of geometry on the integration level. The error between the computed and the reference solution is smaller than 5\% from 4000 DOFs for $C_{1111}$ and less than 1\% for $C_{2211}$.} In this setting, \textcolor{Reviewer3}{ the best resolution is achieved with an accuracy of $0.2$\% in the effective elastic tensor of $C_{1111}$ and the insignificant difference in the off-diagonal entry $C_{2211}$ with $50\,736$ DOFs for 5 finite cells of polynomial order $p=7$}. Additionally, the symmetry and the simplicity of the grid-like meshes used in this embedded approach allow an efficient and flexible application of the boundary condition, independently of the underlying geometry of the RVE.
		
	}
	\subsection{Unit cell with a cubical void: boundary conditions}
	\label{subsec::CubicVoid}
	{
		To study the effect of the boundary conditions on the final homogenized quantities, a unit cell with a centered cubic hole is considered. The size of the void is varied to achieve different porosity states.
		
		The material of the matrix has a Young's modulus $E=200 \, GPa$ and a Poisson's ratio $\nu=0.25$. The cubical RVE has the size of $10 \times 10 \times 10 \,$[mm]. The following numerical parameters are fixed after performing convergence studies: polynomial degree $p=4$ and $8 \times 8 \times 8 $ finite cells (see~\cref{fig::SetupCubicalVoid}).
		Due to a presence of the edge and corner singularities in this geometrical model, a multi-level hp-refinement is performed with the depth towards the singularities $d=3$ \cite{Zander2015, Zander2016, Zander2017a}.
		\begin{figure}[H]
			\centering
			\captionsetup[subfigure]{oneside,margin={0cm,0cm},labelformat=empty}
			\subfloat[(a) Displacmenet field ]{
				\includegraphics[scale=0.15, trim={5cm 0 25cm 6.5cm}, clip]{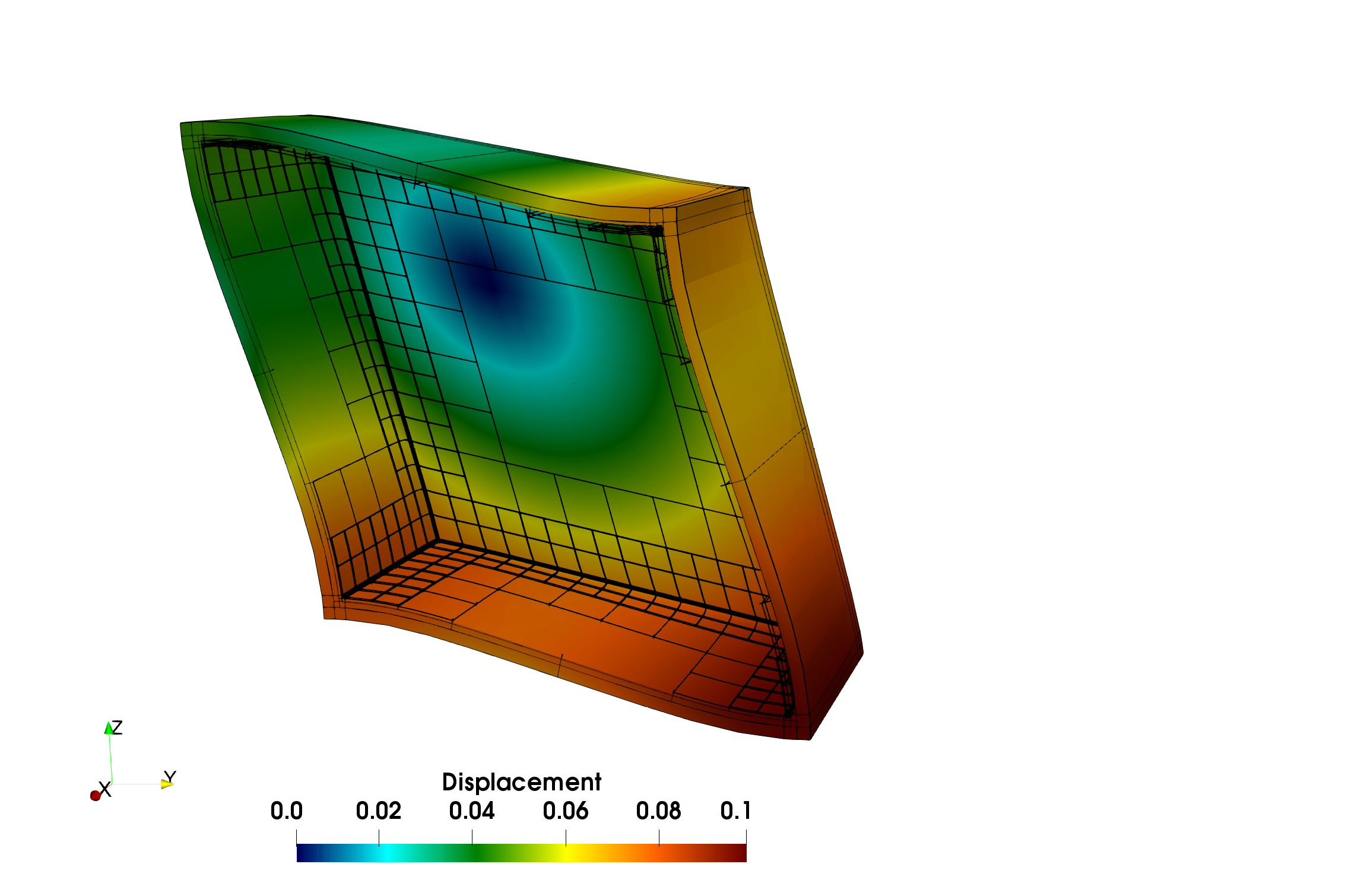} }
			\hspace*{1cm}
			\subfloat[{(b) Von Mises stress field}]{
				\includegraphics[scale=0.15, trim={5cm 0 25cm 4cm}, clip]{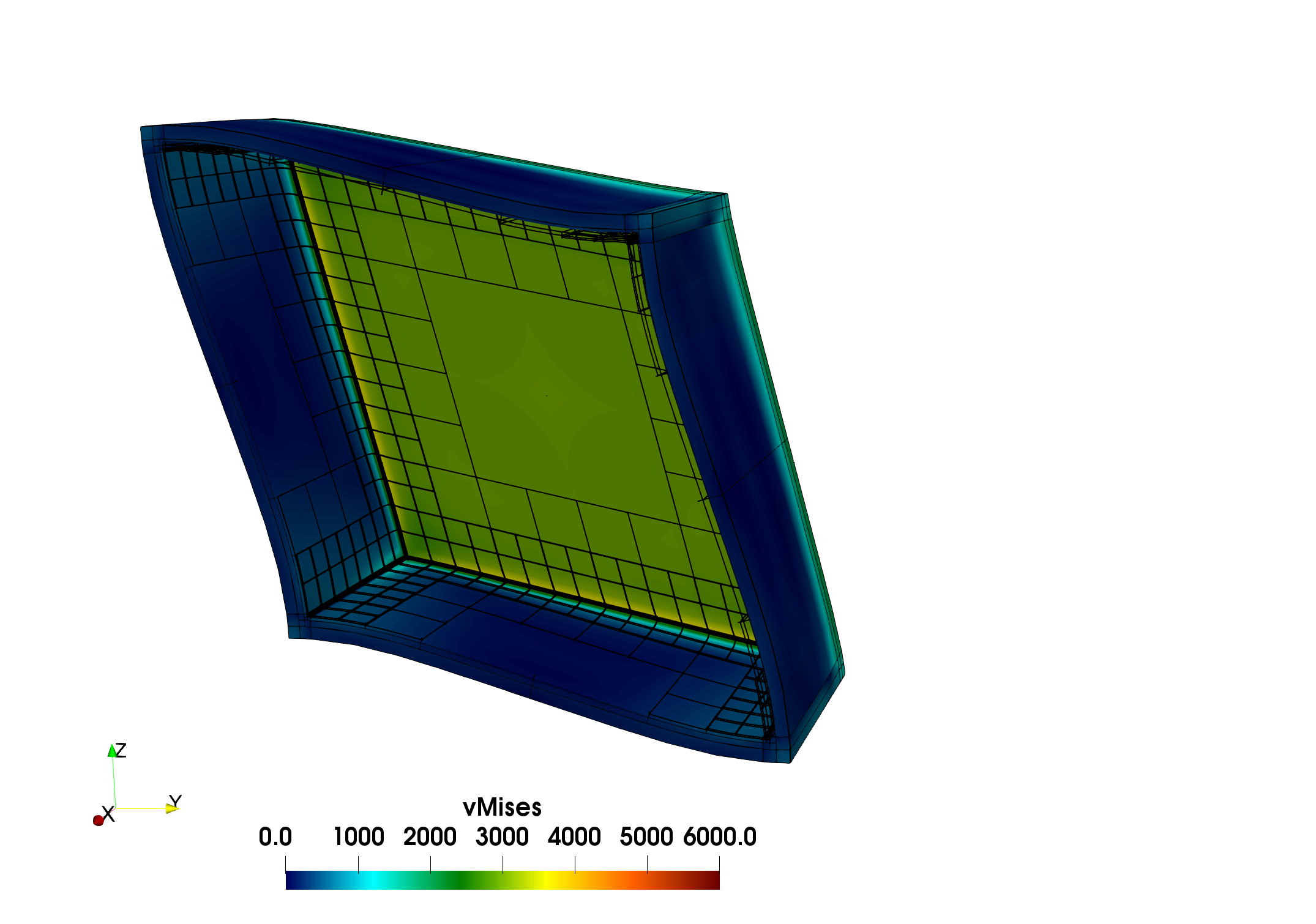} }
			\caption{Deformation patterns of a cubical unit cell with a cubical void of size 9.0[mm] under periodic boundary conditions (scale factor for deformations is 20).}
			\label{fig::SetupCubicalVoid}
		\end{figure}
		
		First, the porosity of the unit cell is varied to study the difference between the boundary conditions described in \cref{subsec::MacroToMicro}. The numerical results are presented together with the analytical Voigt and upper Hashin-Shtrikman bounds. The lower bounds of Hashin-Shtrikman and the analytical Reuss bounds degenerate to zero for porous structures \cite{Gross2017, Willis}. The main assumption of these bounds is a linear isotropic mechanical behavior of the effective domain. However, the numerical simulation of this unit cell did not show a fully isotropic matrix. It follows an isotropic-like structure, but it cannot be expressed in terms of only two constants, e.g. $E$ and $\nu$ or $\lambda$ and $G$. Therefore, \cref{fig::BCsForCubicVoida} shows the shear entry to the elastic tensor $C^*_{1313}$ with respect to the porosity level. The numerical results lie in between the simplest analytical Voigt and Reuss bounds and between the tighter approximation of the Hashin-Shtrikman bounds.	
		
		All boundary conditions follow the order relations in~\cref{eq:BCsAndPBCAsBoundsToEffectiveTensor} for all porosities. Then, the hole size is fixed to 9.0[mm] and the number of the uni cells is increased. As expected, the gap between KUBC and SUBC decreases, while the PBC delivers the same value (see~\cref{fig::BCsForCubicVoidb}).
		\begin{figure}[H]
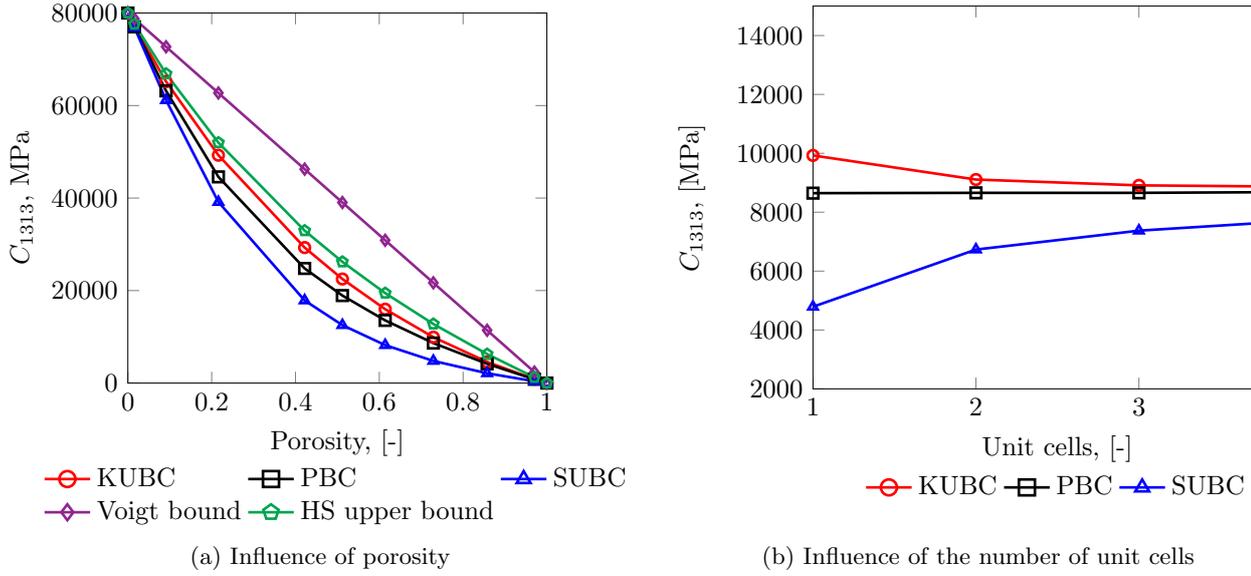

			\centering
			\captionsetup[subfigure]{oneside,margin={0cm,0cm},labelformat=empty}
			\hspace*{-1.2cm}   
			\subfloat[(a) Influence of porosity ]{
				\includegraphics[width=0.5\textwidth,height=0.3\textheight]{\graphDir/TikZ/graphs/windowStudyC1111Porosity.tikz} \label{fig::BCsForCubicVoida}}
			\subfloat[{(b) Influence of the number of unit cells}]{
					\raisebox{2ex}[0pt][0pt]{\def\svgwidth{0.5\linewidth}
					\includegraphics[width=0.5\textwidth,height=0.28\textheight]{\graphDir/TikZ/graphs/UnitCells.tikz}}\label{fig::BCsForCubicVoidb}}
			\caption{Sensitivity study of homogenized properties for a 3D Cubic unit cell with the cubic void to the boundary conditions.}
			\label{fig::BCsForCubicVoid}
		\end{figure}
	}
	    \section{Validation of additive manufacturing product simulation}
\label{sec::AMExamples}
Having all numerical tools at hand, a material characterization workflow is validated on two additively manufactured metal components. \textcolor{Reviewer1}{Specimens 300 and 600 are produced using Inconel\textregistered 718 powder with the standard in-plane square grid cell size of $300 \mu m$ and $600 \mu m$ respectively.}\textcolor{Reviewer2}{ The designed strut size is $96\mu m$.}To this end, CT-scans of final components are segmented using the deep learning segmentation techniques described before. Then, a direct numerical tensile test on full-size specimens is performed. Further, the numerical homogenization is employed and compared against the results of the tensile testing. Finally, the numerical results are compared to the experimental data.

\renewcommand{\graphDir}{./sections/examples/AMSamples/Pictures}
\subsection{Inconel\textregistered 718 Specimen 300}
\label{subsec:CT}
{    
	\Cref{fig:Specimen300} shows a sample of a microporous metallic structure produced by the selective laser melting. The base mechanical properties of Inconel\textregistered 718 are provided by Siemens AG and are $E=190\, GPa$ and $\nu=0.294$. The experimental material properties of these specimens are:
	
	\begin{table}[H]
		\centering
		\begin{tabular}{| c | c | c | c |}
			\hline        
			Specimens & $E_{zz}$, [MPa] & $\nu_{yz}$, [-] & $\nu_{xz}$, [-] \\    \hline
			300 S1-S3 & 18 844--32447 & 0.27--0.35 & 0.23--0.32     \\\hline
		\end{tabular}
		\caption{Specimen 300: Experimental results of a tensile experiment on three specimens.}
		\label{tab::ExperimentalResults}
	\end{table}         
	
	A part of the microstructure of one of the similar specimens was subjected to a computer tomography before its processing to the sample size \textcolor{Reviewer1}{(see an unprocessed cylindrical specimen in ~\cref{fig:CTSlice})}. First of all, the zone of interest for mechanical analysis is determined similar to the one used for the experimental testing. The \textcolor{Reviewer1}{rectangular} overall domain size is $600\times414\times496$ voxels with an interval of $13.08\,\mu m$ in every direction. Due to the high contrast of Hounsfield units, the presence of trapped powder and metal artifacts in the resulting images (see~\cref{fig:CTSlice}), the CT-scan is segmented using the technique described in~\cref{subsec::segmentation}. After running a flood-fill algorithm on the segmentation, a full 3D voxel model is constructed \cref{fig:FullVoxelModel}. \textcolor{Reviewer1}{This specimen corresponds to the middle measurement area of the physical specimen depicted in~\cref{fig:Specimen300}.}
	
	\begin{figure}[H]
		\centering
		\begin{minipage}{.5\textwidth}
			\centering
			\includegraphics[width=.8\textwidth,height=0.2\textheight]{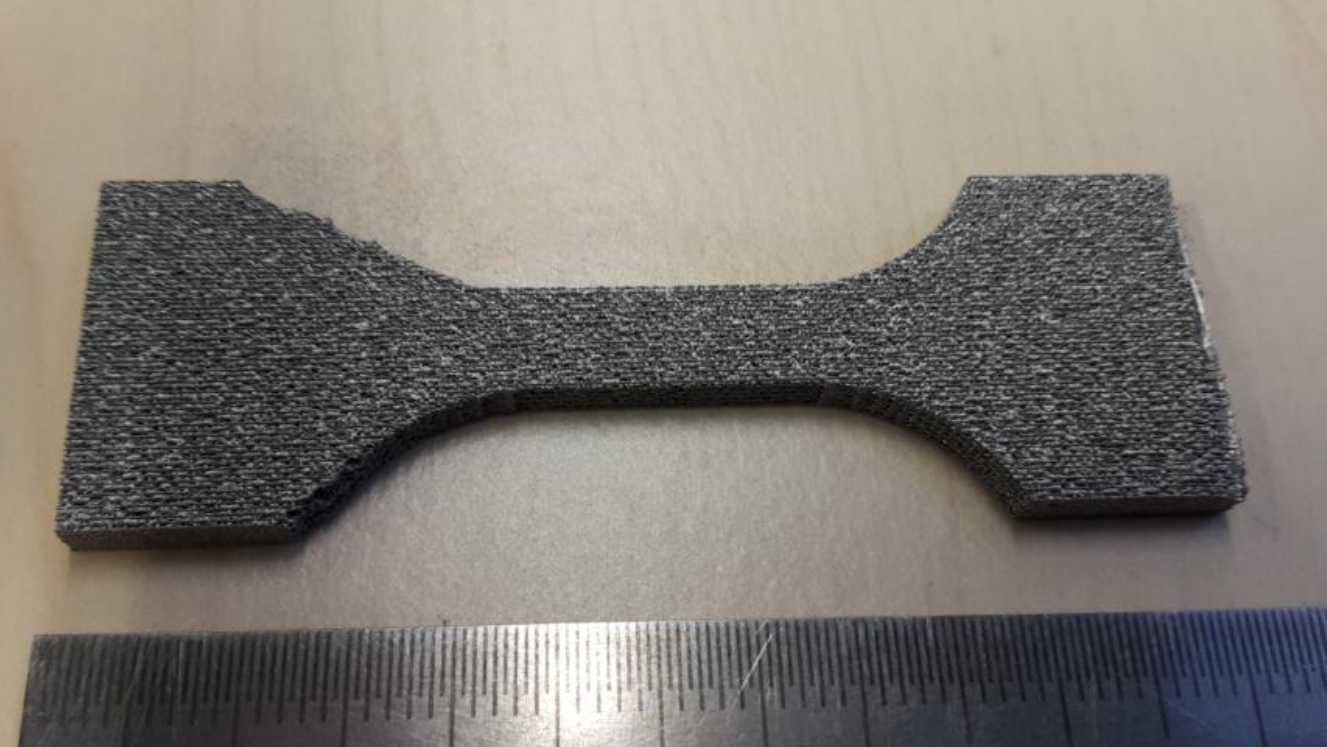}
			\caption{Inconel\textregistered 718 Specimen 300.}    
			\label{fig:Specimen300}
		\end{minipage}%
		\begin{minipage}{.5\textwidth}
			\centering
			\includegraphics[width=.6\textwidth,height=0.2\textheight]{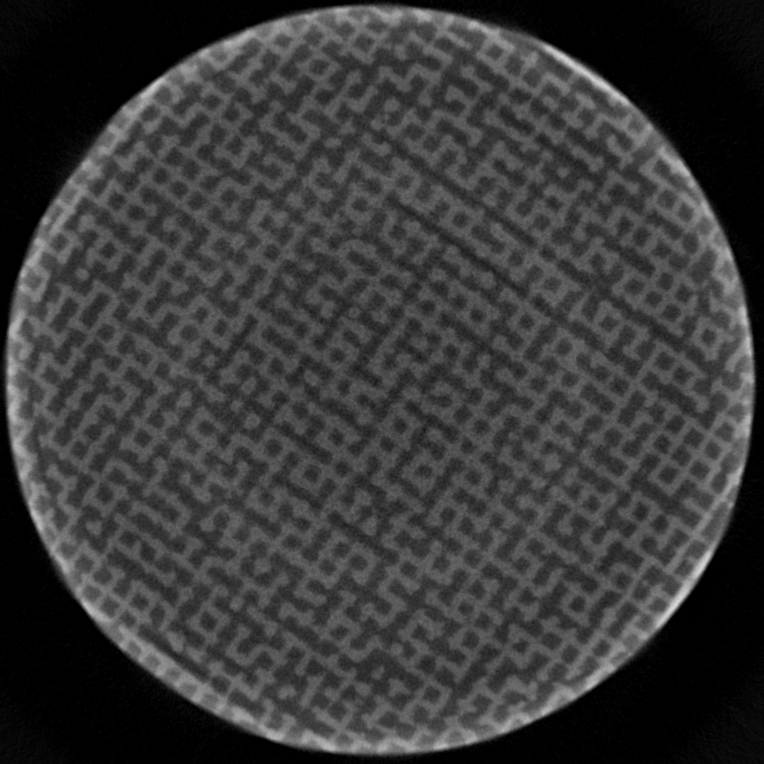}
			\caption{Axial slice of the CT-Scan.}    
			\label{fig:CTSlice}
		\end{minipage}
	\end{figure}
	\begin{figure}[H]
		\centering
		\includegraphics[scale=0.3, trim={0 0cm 0cm 0},clip]{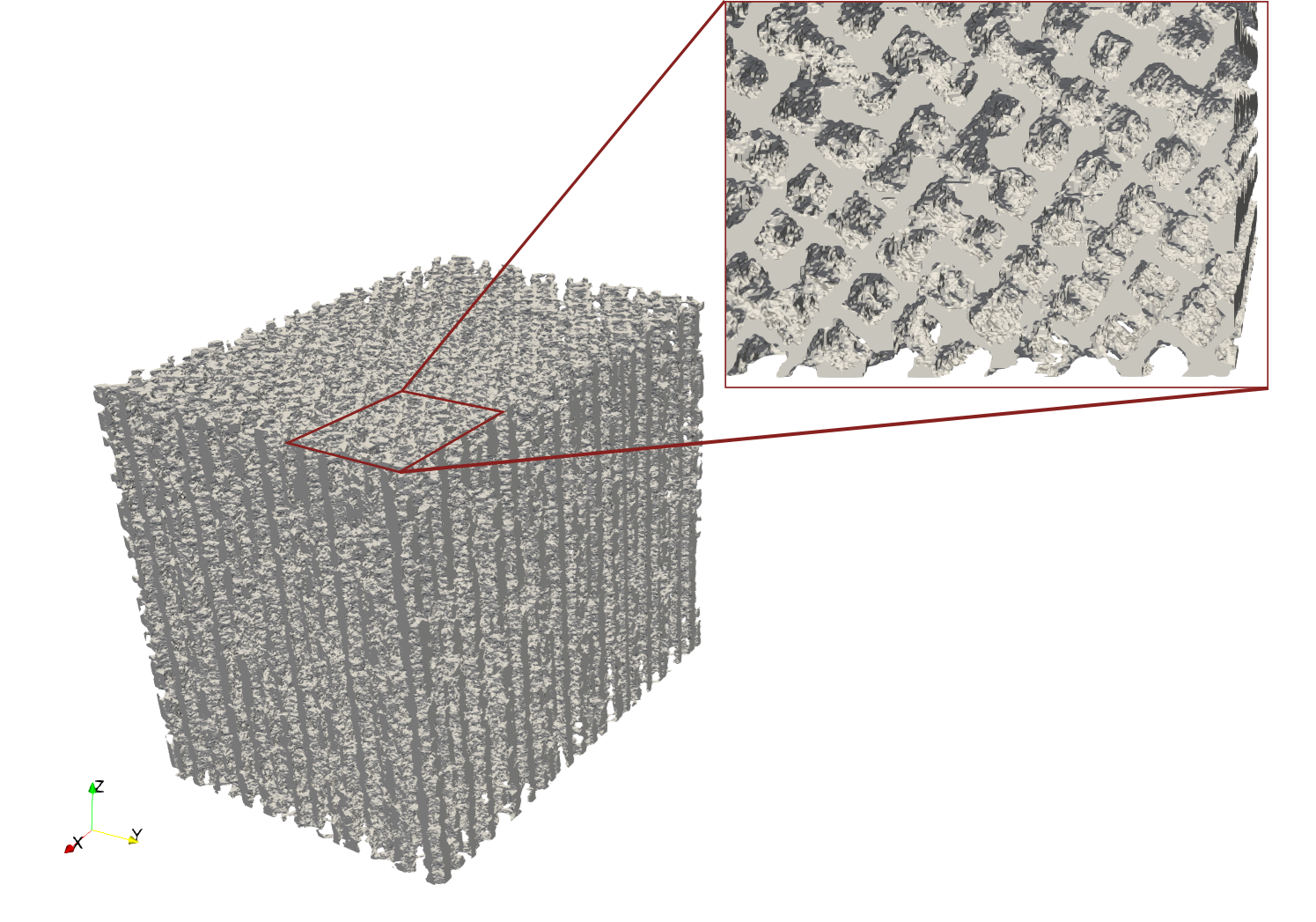}
		\caption{Specimen 300: Voxel model after segmentation.}    
		\label{fig:FullVoxelModel}
	\end{figure}
	
	As it is depicted in~\cref{tab::ExperimentalResults}, the microarchitectured specimens can exhibit large variations in the final mechanical properties. This is due to the difficulty in controlling the process parameters at the microscopic level, which leads to discrepancies in the achieved geometrical details. The analyzed CT-scan does not belong to any of the specimens S1--S3. It belongs to a similar specimen \textcolor{Reviewer1}{S4}, produced with the same process parameters and the same nominal geometry. Therefore, the deviations from the values mentioned in~\cref{tab::ExperimentalResults} can be expected.

	First, an embedded uniaxial numerical tensile experiment (embedded DNS) is performed on a complete \textcolor{Reviewer1}{rectangular specimen as indicated in~\cref{fig:FullVoxelModel}}. A base discretization of $100\times69\times124$ finite cells is used. The polynomial order of the finite cells is elevated uniformly from $p=1$ to $p=5$. According to the voxel-based pre-integration technique described in~\cref{subsec::PreIntegration}, every finite cell consists of $6\times6\times4$ integration partitions. To evaluate the discretization error achieved, the numerical results of an embedded numerical test are compared to the directional module $E_{zz} = 34\,882.85 $ [MPa] of the fully resolved specimen with \textcolor{Reviewer1}{$1\,064\,415\,234$ DOFs} for the high-order Voxel-FEM computation with $p=3$. 
	
	An embedded specimen is computed on the Linux Cluster cluster at the Technical University of Munich, which is equipped with Intel® Xeon® E5-2697 v3 ("Haswell") 60 nodes with 28 cores per node. The computations for $p=1$ up to $p=3$ are performed with hybrid parallelism using 8 nodes with 4 MPI processes on each and 12 OpenMP threads. For a polynomial order $p=4$ the number of nodes is increased to 32, maintaining the number of MPI processes and OpenMP threads and for $p=5$ to 60 nodes. Voxel-FEM computations are performed on the SuperMUC cluster on 150 nodes for $p=1$, 600 nodes for $p=2$ and on 500 nodes for the $p=3$.
	
	\renewcommand{\graphDir}{./sections/examples/AMSamples/Pictures/TikZ/FullTensileTest/graphs}
	\newcommand{\dataDir}{./sections/examples/AMSamples/Pictures/TikZ/FullTensileTest/data}
	\begin{figure}[H]
		\centering
		\hspace*{-1.5cm}
		\begin{minipage}{.5\textwidth}
			\centering
			\includegraphics[height=0.4\textheight]{\graphDir/convergenceFull.tikz}
			\caption{Convergence of the directional Young's modulus $E_{zz}$.}    
			\label{fig:ConvergenceDNS}
		\end{minipage}        \renewcommand{\graphDir}{./sections/examples/AMSamples/Pictures/TikZ/HomCTScan/graphs}
		\renewcommand{\dataDir}{./sections/examples/AMSamples/Pictures/TikZ/HomCTScan/data}
		\begin{minipage}{.5\textwidth}
			\vspace*{-0.25cm}
			\centering
			\includegraphics[height=0.42\textheight]{\graphDir/stressStrain.tikz}
			\caption{Stress-strain curve for Specimen 300.}    
			\label{fig::HomogenizationInconel300}
		\end{minipage}
	\end{figure}

	An example of a resulting displacement field and a von Mises stress state is shown in \cref{fig:CTScanDisplacementStress}. The convergence curves (see~\cref{fig:ConvergenceDNS}) show that the last embedded computation with the polynomial degree of $p=5$ leads to 11.3\% error relative to an overkill solution provided by Voxel-FEM. Therefore, this discretization is kept for further numerical tests.
	
	\renewcommand{\graphDir}{./sections/examples/AMSamples/Pictures}
	\begin{figure}[H]
		\captionsetup[subfigure]{labelformat=empty}
		\centering
		\subfloat[]{\includegraphics[scale=0.25, trim={8cm 0cm 8cm 0cm},clip]{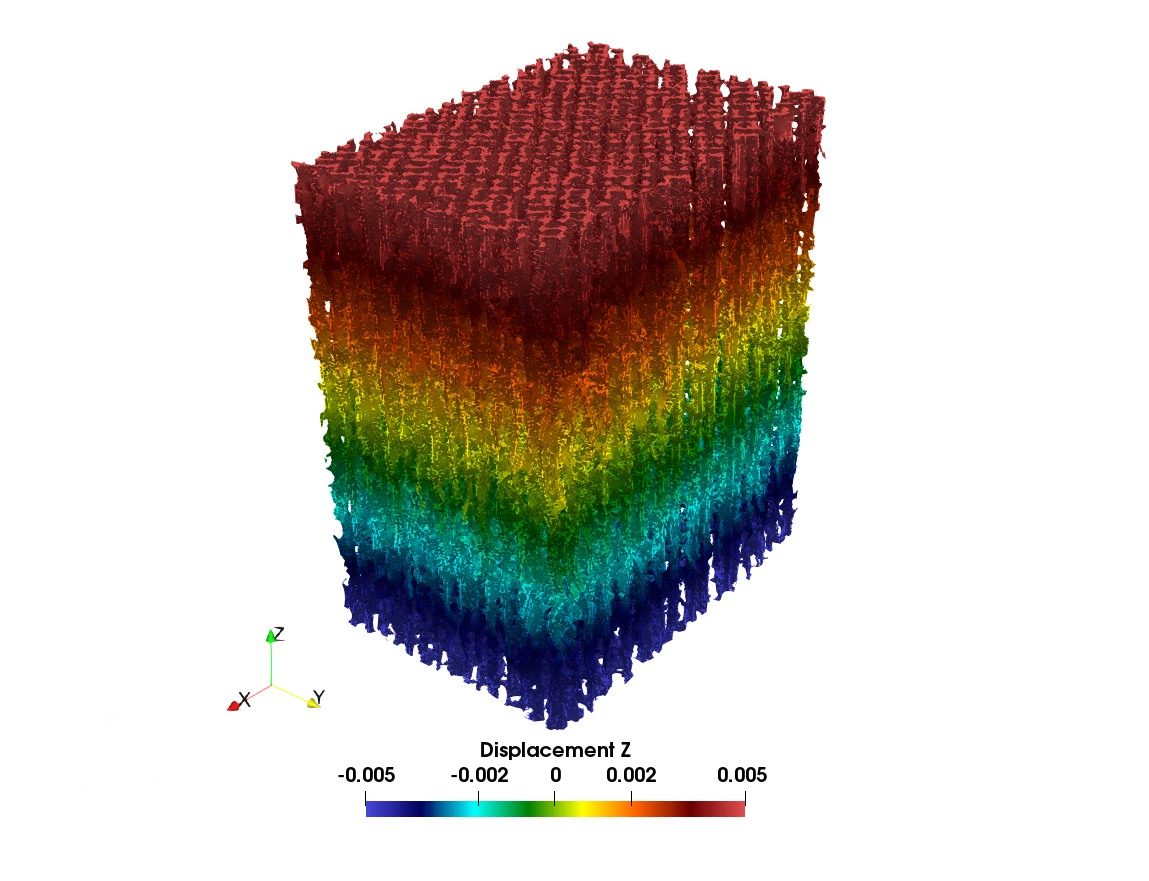}}
		\subfloat[]{\includegraphics[scale=0.25, trim={8cm 0cm 8cm 0cm},clip]{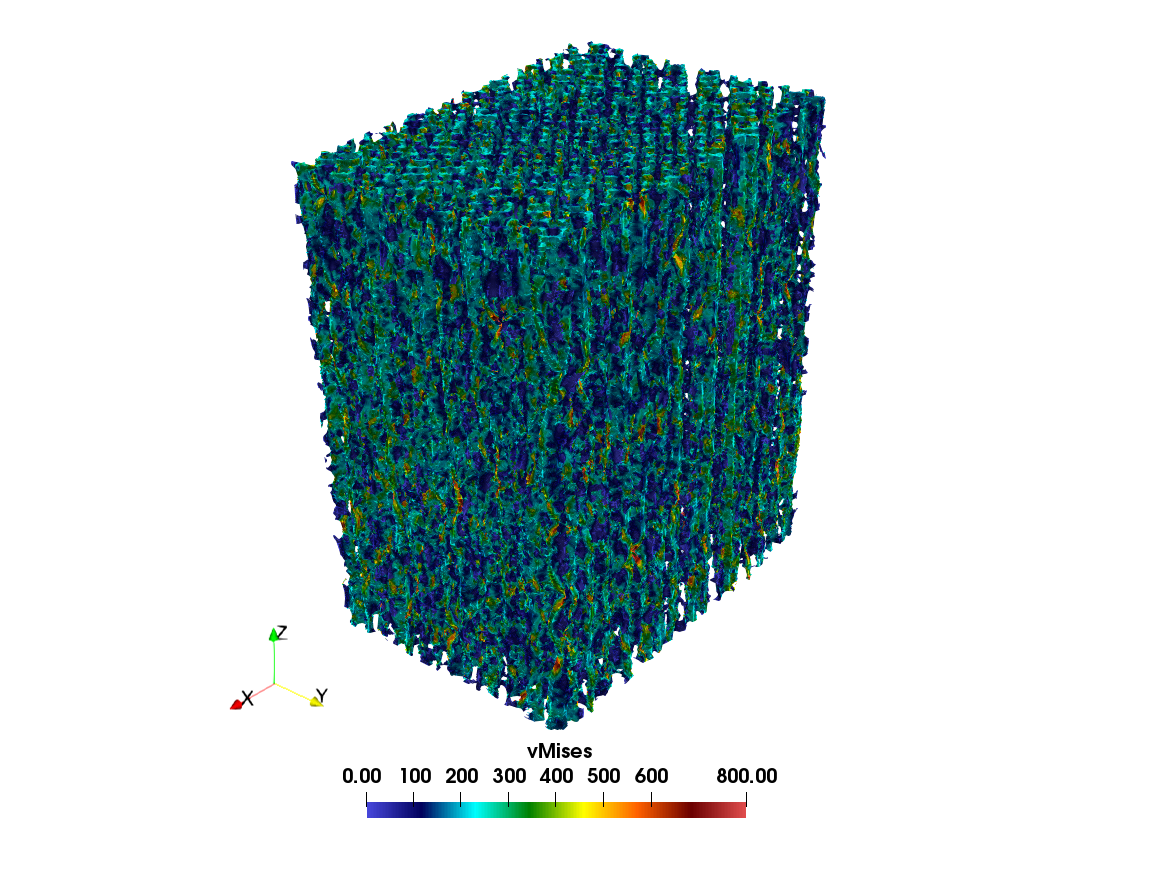}}
		\caption{Displacement and von Mises stress fields of Specimen 300 under uniaxial tension in the deformed state.}
		\label{fig:CTScanDisplacementStress}
	\end{figure}
	
	Next, the homogenization techniques described in \cref{sec::numericalHomogenization} are used and verified against the DNS computation. Homogenization is performed through the entire CT-scan by moving an RVE window. 168 RVEs are computed to evaluate the variation of the apparent properties concerning the location. \textcolor{Reviewer1}{ A coefficient of variation (CV) is computed to quantify this difference as follows:
	\begin{equation}
	CV=\frac{\sigma}{\mu}
	\end{equation} 
where $\sigma$ is a standard deviation of the homogenized Young's modulus $E$ in a statistical set of 168 homogenized values and $\mu$ is a mean value of the quantity of interest $E$ over the whole set. } Considering the microstructure of this volume, the size of the RVE is chosen as $84\times84\times80$ voxels. For a fair comparison between numerical homogenization and numerical tensile test, the same discretization should be kept for both cases.  This means that every RVE is discretized with $14\times14\times20$ finite cells of polynomial order $p=5$ with $6\times6\times4$ integration partitions in each. The resulting stress-strain curves are shown in~\cref{fig::HomogenizationInconel300}. The quantitative comparison is provided in~\cref{tab::QuantitativeComparisonInconel300}.
	
	\begin{table}[H]
		\centering
		\begin{tabular}{| c | c | c |}
			\hline        
			Test & $E_{mean}$, [MPa] & CV, [\%] \\    \hline
			Experiment lower &    18 884 &  - \\\hline
			Experiment upper &     32 447 &  -     \\\hline
			DNS( embedded, $p=5$ ) &     39 351.31 &- \\\hline
			KUBC &     44 438.52&   12.06 \\\hline
			PBC &     38 417.80 &   15.50 \\\hline
		\end{tabular}
		\caption{Quantitive comparison of the numerical results for the homogenized Young's modulus $E_{zz}$ and its coefficient of variation CV.}
		\label{tab::QuantitativeComparisonInconel300}
	\end{table}         
	\Cref{tab::QuantitativeComparisonInconel300} shows that the coefficient of variation of the computed homogenized Young's modulus is about 15\%. Therefore, the chosen volume is considered as representative for the purposes of this application (\cite{Fan}). The KUBC, as expected, are delivering an upper bound. SUBC are not applicable due to the high concentration of the voids randomly crossing the boundary of the RVEs. The Periodic Boundary Conditions delivered the best estimate, deviating from the numerical tensile experiment by $2.4\%$. 
	
	As expected, the numerical results differ from the experimental values. The numerical analysis is held \textcolor{Reviewer1}{on the same "as-planned" geometrical specimen as the ones mentioned in~\cref{tab::ExperimentalResults}, but not the ones mentioned as S1--S3}. This proves the importance of the induced geometrical defects and shows that the same intended geometry under the same process parameters can lead to a different microstructure. Despite the discrepancies between numerical and experimental results, the accurate agreement between the effective Young's modulus determined by the tensile test and the homogenization procedure gives confidence in the \textcolor{Reviewer1}{proposed numerical approaches. } \textcolor{Reviewer3}{Furthermore the variation of the experimentally determined mechanical properties is large due to the presence of process induced defects. Future work will, therefore, concentrate on generating statistically equivalent samples. The proposed numerical approach then provides a computational tool to compute the variation of the mechanical properties due to such geometrical and topological defects.}
	
	An important aspect to be mentioned is the numerical cost for both numerical approaches to describe the material behavior of such structures. For the embedded DNS simulation with $p=5$ and with $6\times6\times4$ integration partitions the wall run-time on 60 nodes is ca. 48 min. The integration of the stiffness matrix took ca. 7 s. These computational times highlight the efficiency of the used pre-integration technique and emphasize the advantage of the high-order embedded approach. Nevertheless, at least 6 numerical tests would be required for a complete material characterization, leading to an increased need in parallel resources. By contrast, the numerical homogenization is embarrassingly parallel. The RVEs at different locations can be run independently of each other without any communication needed. The full numerical homogenization procedure with PBC and $p=5$ took ca. 5 min on average on the Intel Core i7-4790 CPU with 4 OpenMP threads. This computation delivers a full material tensor and does not need to be repeated. Even if the complete analysis is desired, i.e. all existing RVEs are considered, the computational cost required is much smaller than that needed to perform numerical testing of the full specimen.

}

\subsection{Inconel\textregistered 718 Specimen 600}
\label{subsec::Inconel600}
{
	In this subsection, a sample 600 of Inconel\textregistered 718 is analyzed (see~\cref{fig::Specimen600}). Due to a larger space between individual laser tracks, there is no powder trapped within the specimen. Moreover, the process parameters for this setup are easier to control. Therefore, the microstructure does not present a large variety of defects and is more reproducible. Nevertheless, the metal artifacts still remain present in the CT-Scan (see ~\cref{fig::CTSlice600}). The deep learning segmentation technique was applied to \textcolor{Reviewer1}{generate a} reliable 3D voxel model depicted in~\cref{fig::FullVoxelModel600}
	
	\begin{figure}[H]
		\centering
		\begin{minipage}{.5\textwidth}
			\centering
			\vspace*{1.4cm}
			\includegraphics[scale=0.44]{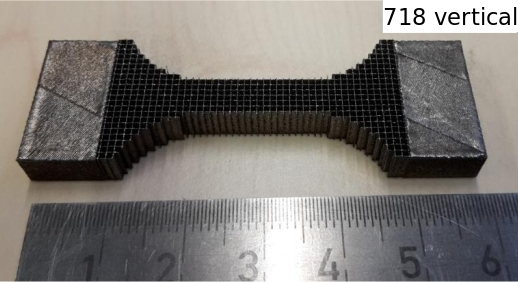}
			\caption{Inconel\textregistered 718 Specimen 600.}    
			\label{fig::Specimen600}
		\end{minipage}%
		\begin{minipage}{.5\textwidth}
			\centering
			\includegraphics[scale=0.3, trim={0cm 0cm 25cm 3cm},clip]{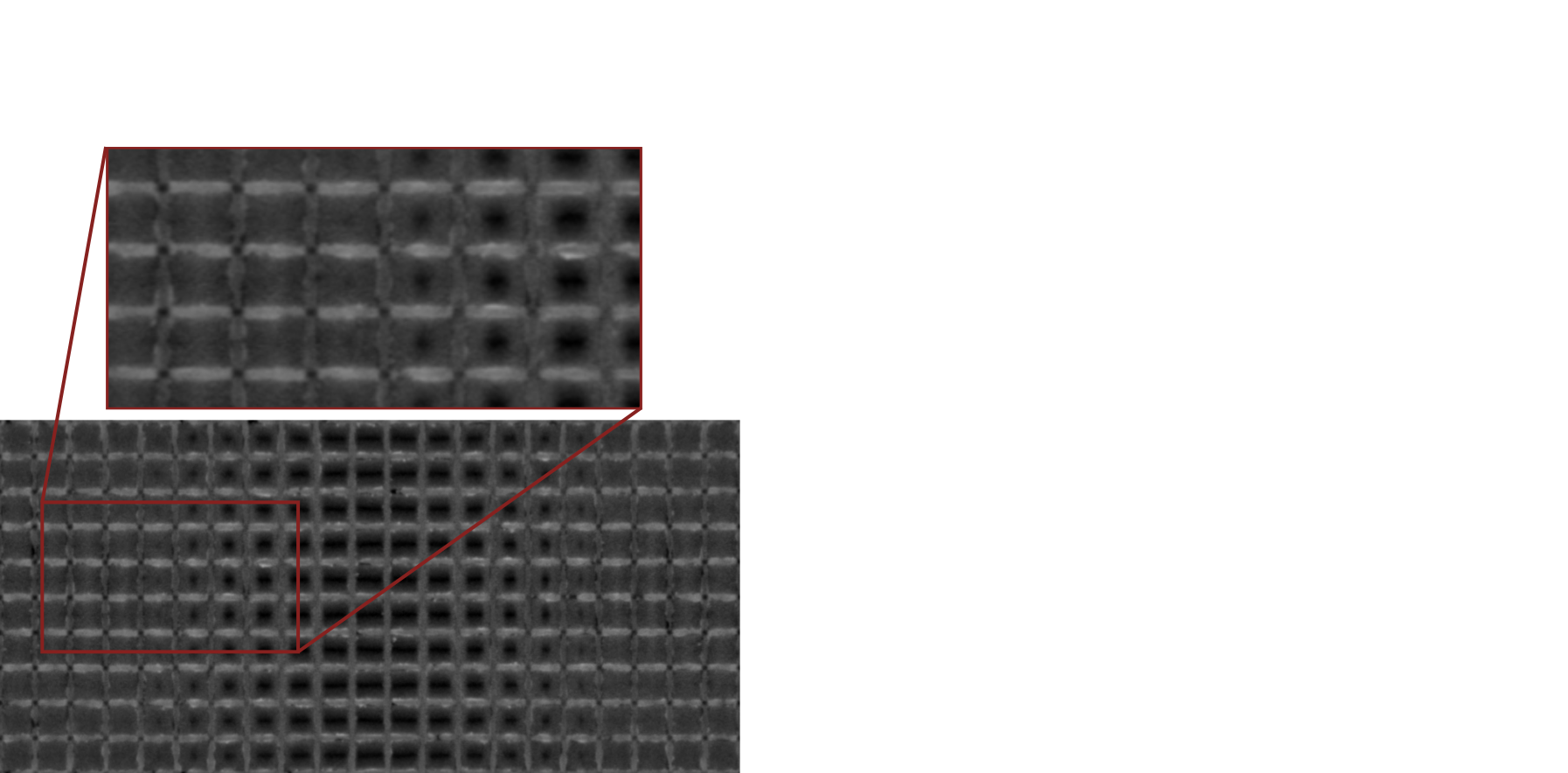}
			\caption{Coronal slice of the CT-Scan for Specimen 600.}    
			\label{fig::CTSlice600}
		\end{minipage}
	\end{figure}  
	\begin{figure}[H]
		\centering
		\includegraphics[scale=0.25, trim={0cm 0cm 0cm 0cm},clip]{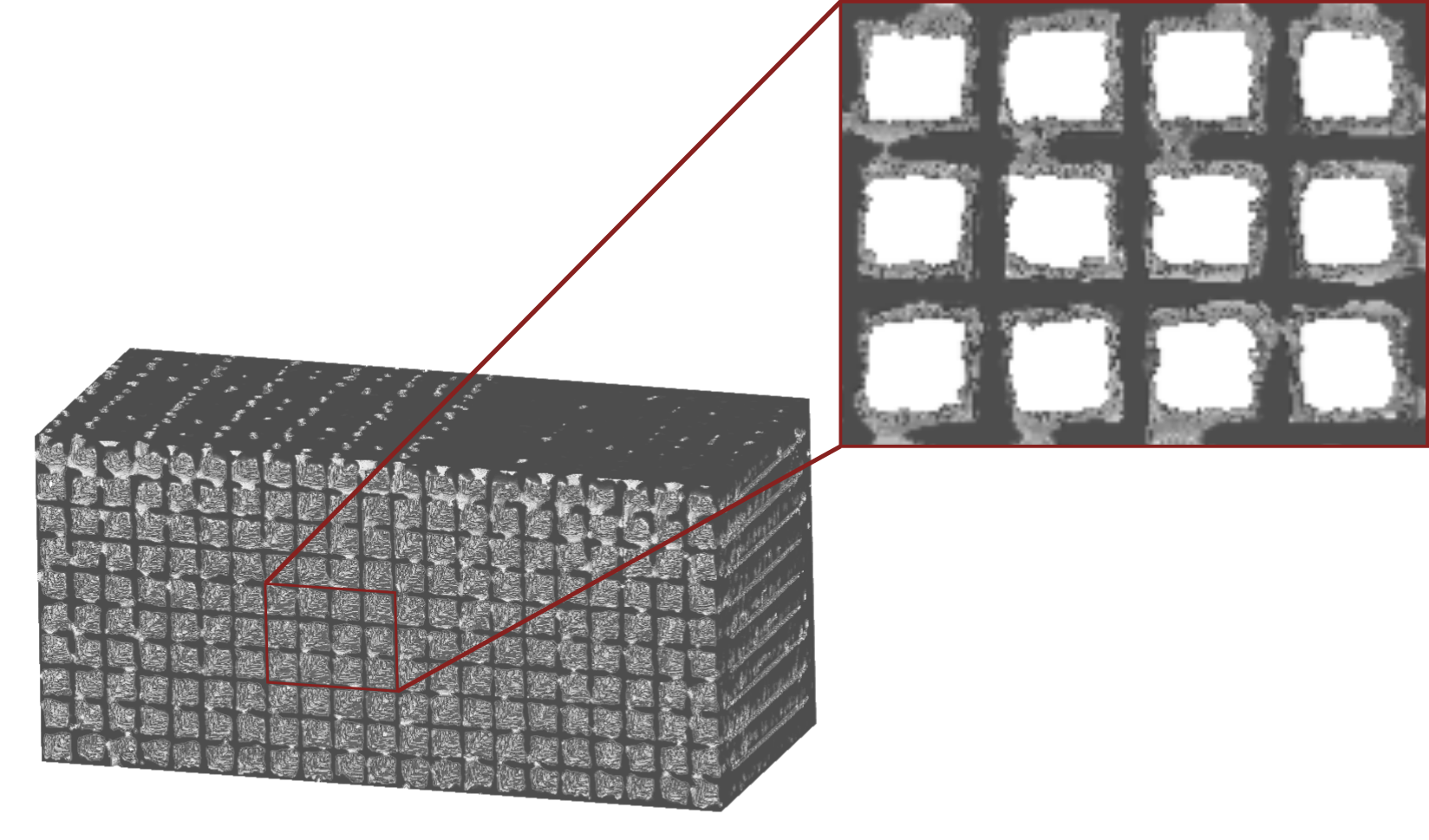}
		\caption{Specimen 600: Voxel model after segmentation.}    
		\label{fig::FullVoxelModel600}
	\end{figure}
	
	In this case, the provided CT-scan corresponds to the specimen L2 and is taken before the elastic testing. The experimental results for 3 specimens are summarized in~\cref{tab::ExperimentalResults600}. As the experiments are performed with the help of optical microscopy, \textcolor{Reviewer1}{two orthogonal cameras are installed tracking the lateral extension. One camera is frontal to the specimen, tracking the extension of the wide side of the middle of the specimen. While the other camera is tracking the extension from the narrow side of the texting specimen. }
	\begin{table}[H]
		\centering
		\begin{tabular}{| c | c | c | c |}
			\hline        
			Specimens & $E_{xx}$, [MPa] & $\nu_{xy}$, [-] & $\nu_{xz}$, [-] \\    \hline
			600 L1-L3 & 15 339...26 731 & -0.04..0.13 & 0.05..0.14     \\\hline
			600 L2 narrow & 20 851 & -0.04 & 0.08    \\\hline
			600 L2 wide   & 25 915 & - & -    \\\hline
		\end{tabular}
		\caption{Specimen 600: Experimental results of a tensile experiment on three specimens.}
		\label{tab::ExperimentalResults600}
	\end{table}        
	
	The same workflow as before is followed. The total number of voxels in this specimen is $800\times368\times400$. Two different discretizations for the embedded high-order FCM simulations (embedded DNS) are studied. The first one has $100\times46\times100$ finite cells, and the second one consists of $200\times92\times200$ cells. Polynomial degree for each case is raised from $p=1$ to $p=6$. Again, the embedded simulations are compared to a high-order Voxel-FEM solution with $E_{xx} = 23\,584.62 $ [MPa] of a fully resolved specimen with $1\,428\,159\,774$ DOFs for $p=4$. \Cref{fig::ConvergenceDNS600} summarizes the relative error in the Young's modulus $E_{xx}$ for the two discretizations. The number of nodes and processes is the same as for specimen 300. The first discretization is considered acceptable as it provides a $9.74\%$ error with respect to the overkill solution.
	
	\renewcommand{\graphDir}{./sections/examples/AMSamples/Pictures/TikZ/FullTensileTestNewMay/graphs}
	\newcommand{\dataDir}{./sections/examples/AMSamples/Pictures/TikZ/FullTensileTestNewMay/data}
	\begin{figure}[H]
		\centering
		\hspace*{-0.5cm}
		\begin{minipage}{.5\textwidth}
			\centering
			\vspace*{-0.01cm}
			\includegraphics[height=0.37\textheight]{\graphDir/convergenceFull.tikz}
			\caption{Specimen 600: Convergence of the directional Young's modulus $E_{xx}$.}    
			\label{fig::ConvergenceDNS600}
		\end{minipage}        
		\renewcommand{\graphDir}{./sections/examples/AMSamples/Pictures/TikZ/HomCTScanNew/graphs}
		\renewcommand{\dataDir}{./sections/examples/AMSamples/Pictures/TikZ/HomCTScanNew/data}
		\hspace*{-0.5cm}
		\begin{minipage}{.5\textwidth}
			\centering
			\vspace*{-0.4cm}
			\includegraphics[height=0.35\textheight]{\graphDir/stressStrain.tikz}
			\vspace*{0.4cm}
			\caption{Stress-strain curve for Specimen 600.}    
			\label{fig::HomogenizationInconel600}
		\end{minipage}
	\end{figure}
	
	To evaluate the accuracy of the homogenization techniques presented in~\cref{sec::numericalHomogenization} the same model is analyzed by means  of the numerical homogenization techniques. The size of the RVE is chosen to be $80\times80\times80$ voxels, discretized with $10 \times20\times10$ finite cells of $p=6$. The stress-strain curves are shown in~\cref{fig::HomogenizationInconel600}. Quantitative results are summarized in~\cref{tab::QuantitativeComparisonInconel600}.
	
	\begin{table}[H]
		\centering
		\begin{tabular}{| c | c | c |}
			\hline        
			Test & $E_{mean}$, [MPa] & CV, [\%] \\    \hline
			Experiment &     23 383 &  3 580.79     \\\hline
			DNS( embedded, $p=6$ ) &     25 881.40 &   - \\\hline
			KUBC &     29 711.98 &  10.51 \\\hline
			PBC &  27 176.42 &   12.20 \\\hline
		\end{tabular}
		\caption{Quantitative comparison of the numerical results for the homogenized Young's modulus $E_{xx}$ and its coefficient of variation CV.}
		\label{tab::QuantitativeComparisonInconel600}
	\end{table}    
	
	The Periodic Boundary Conditions provide a slightly stiffer response but agree with the numerical tensile test with an error of 5\%. It is noteworthy that the results of this simulation are in an exceptional agreement with the experimental tests. These findings suggest that the discrepancies between the numerical and experimental tests for the previously described Specimen 300 arise from the expected geometrical difference in the printed structure.
	
	The numerical costs of these simulations are similar to the ones for Specimen 300. The direct numerical tensile test for $p=6$ is performed on 80 nodes of the SuperMUC cluster with 4 MPI processes on each. The total wall clock time is ca. 57 min, where the integration of the global stiffness matrix takes ca. 3 s. The numerical homogenization of one RVE with PBC is approximately 2 min on average on the Intel Xeon E5-2690 CPU with 16 OpenMP threads.
	
	While these investigations focused on the verification and validation of the proposed numerical workflow, further work is required to develop a methodology to be able to characterize the observed variation of the material parameters.
}    

}
	\section{Summary and Outlook} 
\label{sec:Conclusion}
 {  
  In this paper, an alternative image-based material characterization workflow for AM microarchitectured structures is proposed. This approach is based on the embedded high-order Finite Cell Method, which eliminates the need for a tedious mesh generation for imperfect microstructures and allows a direct numerical analysis on the CT-images of final products with a reduced amount of necessary Degrees of Freedom. However, as the materials under consideration are metals, diverse artifacts known for this group of materials were observed in the computer tomographic images. To this end, an advanced deep learning segmentation was integrated into the workflow. It facilitates the detection of the metal artifacts and provides a reliable computational domain resulting in a more accurate material characterization. Direct numerical testing and a numerical homogenization technique were considered as two possible ways to characterize the material behavior of AM components. For a direct numerical tensile experiment, an efficient parallel implementation of the embedded method together with the voxel-based pre-integration technique was introduced. This allowed computations on complete AM testing specimens resolving the microstructural details. However, as a cluster is usually needed for direct numerical computations, numerical homogenization was introduced as an alternative approach to characterize linear elastic behavior of AM final parts. In this context, Cartesian grid meshes used in the FCM were advantageous to apply Periodic Boundary Conditions and to analyze porous non-periodic domains.  Both numerical approaches showed an excellent agreement with each other. Yet, when comparing the numerical results to the experiments, it was demonstrated that such a good agreement can be achieved only when an exact geometrical description of the analyzed specimen is available. 

While this study is the first step towards a complete material characterization of AM products with the process-induced defects, future research to enhance this concept should be undertaken. Future work will, first, investigate a possible way to generate statistically similar structures to support a large variation in the experimental results among similarly produced specimens. Secondly, an extension of these concepts to the non-linear mechanical behavior will be addressed. Finally, the numerical behavior of these geometries under more complex loading will be examined.            
  }

\section*{Acknowledgements} 
We gratefully acknowledge the support of Siemens AG sponsoring this research and the support of Deutsche Forschungsgemeinschaft (DFG) through TUM International Graduate School of Science and Engineering (IGSSE), GSC 81. This work was also supported by the Bayerische Kompetenznetzwerk f\"{u}r Technisch-Wissenschaftliches Hoch- und Höchstleistungsrechnen (KONWIHR). The authors also gratefully acknowledge the Gauss Centre for Supercomputing e.V. (www.gauss-centre.eu) for funding this project by providing computing time on the GCS Supercomputer SuperMUC and on the Linux Cluster CoolMUC-2 at Leibniz Supercomputing Centre (www.lrz.de).

\newpage
\bibliographystyle{apalike}
 \bibliography{library}

\end{document}